\documentclass[a4paper,10pt,twocolumn]{IEEEtran}

\usepackage[utf8]{inputenc}
\usepackage[cmex10]{amsmath}
\usepackage{graphicx}
\usepackage{amssymb}
\usepackage{amsthm}
\usepackage{subfigure}
\usepackage{cite}
\usepackage{color}
\usepackage{pifont}
\newcommand{\cmark}{\ding{51}}
\newcommand{\xmark}{\ding{55}}

\theoremstyle{remark}

\usepackage[ruled,norelsize]{algorithm2e}
\makeatletter
\newcommand{\removelatexerror}{\let\@latex@error\@gobble}
\makeatother

\usepackage{algorithmicx}

\newcommand{\col}{black}

\begin{document}

\title{Millimeter Wave MIMO Channel Estimation using Overlapped Beam Patterns and Rate Adaptation}

\author{
    {Matthew Kokshoorn, \textit{Graduate Student Member, IEEE}, He Chen, \textit{Member, IEEE}, Peng Wang, \textit{Member, IEEE}, Yonghui Li, \textit{Senior Member, IEEE}, and Branka Vucetic, \textit{Fellow, IEEE}} 
\thanks{
The material in this paper was presented in part at the IEEE International Conference on Communications, London, England, June 2015 \cite{Kokshoorn} and will be presented in part at IEEE Global Communication Conference, Washington D.C., United States, December 2016 \cite{Koks1612}. 

This research was supported by an Australian Postgraduate Award (APA) and ARC grants DP150104019 and FT120100487. The research was also supported by funding from the Faculty of Engineering and Information Technologies, The University of Sydney, under the Faculty Research Cluster Program and the Faculty Early Career Researcher Scheme.

M. Kokshoorn, H. Chen, Y. Li, and B. Vucetic are with School of Electrical and Information Engineering, The University of Sydney, Sydney, NSW 2006, Australia (email: \{matthew.kokshoorn, he.chen, yonghui.li, branka.vucetic\}@sydney.edu.au).

P. Wang was with the School of Electrical and Information Engineering, The University of Sydney, Sydney, NSW 2006, Australia.  He is now with Huawei Technologies Sweden AB, Kista 164 40, Sweden (email: wp\_ady@hotmail.com).}}

\maketitle

\begin{abstract}
This paper is concerned with the channel estimation problem in Millimeter wave (mmWave) wireless systems with large antenna arrays. By exploiting the inherent sparse nature of the mmWave channel, we first propose a fast channel estimation (FCE) algorithm based on a novel overlapped beam pattern design, which can increase the amount of information carried by each channel measurement and thus reduce the required channel estimation time compared to the existing non-overlapped designs. We develop a maximum likelihood (ML) estimator to optimally extract the path information from the channel measurements. Then, we propose a novel rate-adaptive channel estimation (RACE) algorithm, which can dynamically adjust the number of channel measurements based on the expected probability of estimation error (PEE). The performance of both proposed algorithms is analyzed. For the FCE algorithm, an approximate closed-form expression for the PEE is derived. For the RACE algorithm, a lower bound for the minimum signal energy-to-noise ratio required for a given number of channel measurements is developed based on the Shannon-Hartley theorem. Simulation results show that the FCE algorithm significantly reduces the number of channel estimation measurements compared to the existing algorithms using non-overlapped beam patterns. By adopting the RACE algorithm, we can achieve up to a 6dB gain in signal energy-to-noise ratio for the same PEE compared to the existing algorithms.
\end{abstract}

\section{Introduction}
Millimeter wave (mmWave) communication has been shown to be a promising technique for next generation wireless systems due to a large expanse of available spectrum \cite{pi2011introduction,rappaport2013millimeter,rheath,hong2014study}. This spectrum, ranging from 30GHz to 300GHz, offers a potential 200 times the bandwidth currently allocated in today's mobile systems \cite{Rappaport200}. However, a critical challenge in exploiting the mmWave frequency band is its severe signal propagation loss compared to that over conventional microwave frequencies \cite{rheath,zhang2010channel,torkildson2010channel}. To compensate for such a loss, large antenna arrays can be employed to achieve a high power gain. Fortunately, owing to the small wavelength of mmWave signals, these arrays can be packed into small areas at the transmitter and receiver \cite{hur2013millimeter,biglarbegian2011optimized}. For such mmWave systems, channel state information (CSI) is essential for effective communication and precoder design. However, the use of large antenna arrays results in a large multiple-input multiple-output (MIMO) channel matrix. This makes the channel estimation of mmWave systems very challenging due to the large number of channel parameters to be estimated. Moreover, owing to the high frequency, it is often not feasible to obtain digital samples from each antenna \cite{pi2011introduction}. To resolve this high frequency sampling problem, analog beamforming techniques have been proposed and widely adopted in open literature (see \cite{5284444,beamcodebook,chen2011multi,hur2013millimeter} and references therein). The main idea of analog beamforming is to control the phase of the signal transmitted or received by each antenna via a network of analog phase shifters.

Using analog beamforming techniques, the most straightforward channel estimation method is to exhaustively search in all possible angular directions. Specifically, consider a system with $N$ transmit antennas and $N$ receive antennas. If we aim at achieving a minimum angular resolution of $\pi/N$, an exhaustive search-based channel estimation would then require a set of $N$ transmit beamforming vectors at the transmitter designed to span all possible beam directions and likewise with $N$ receive beamforming vectors at the receiver. By searching all possible $N^2$ combinations, an $N \times N$ matrix can be formed whose entries represent the channel gains between the $N$ transmit and the $N$ receive beams. This matrix is commonly referred to as the virtual channel matrix \cite{sayeed2002deconstructing,hong2003spatial,lagarias1998convergence,Jianhua,Seo}.
Despite the large number of entries expected for the mmWave MIMO channel matrix, it has been shown in recent measurements \cite{rappaportMeasure,Akdeniz} that the mmWave channel exhibits sparse propagation characteristics in the angular domain. That is, there are only a few dominant propagation paths in mmWave channels. This sparsity can be seen in the virtual channel matrix, as only a limited number of transmit and receive direction combinations have a non-zero gain \cite{sayeed2002deconstructing}. Therefore, the key objective of mmWave channel estimation is to identify these paths so that the transceiver can align the transmit and receive beams along these paths.

Recently, some compressed sensing based channel estimation algorithms have been proposed to explore the channel sparsity in mmWave systems, e.g., \cite{Kokshoorn,rheath,hur2013millimeter,Compressed_Channel_Sensing}. The fundamental idea in some of these approaches is to search multiple transmit/receive directions in each measurement, by creating initial beam patterns that span a wider angular range than those used by the exhaustive search. Similar adaptive beamforming algorithms and multi-stage codebooks were proposed in \cite{IEEE_standard,chen2011multi,Tsang,Zhang_Kung}. {\color{\col}More recent work \cite{xiao2016hierarchical} has also shown that such hierarchical codebooks can be achieved with a single RF chain by exploiting sub-array and deactivation (turning-off) antenna processing techniques.} By initially using wider beam patterns, multi-stage approaches are able to reduce the number of measurements required for channel estimation. However, this introduces a loss of directionality gain, leading to a lower signal-to-noise ratio (SNR) at the receiver and a higher probability of estimation error (PEE). In this sense, there exists a challenging trade-off between estimation time and accuracy for mmWave channel estimation.

\begin{figure}[!t]
\centering
\includegraphics[width=3.5in]{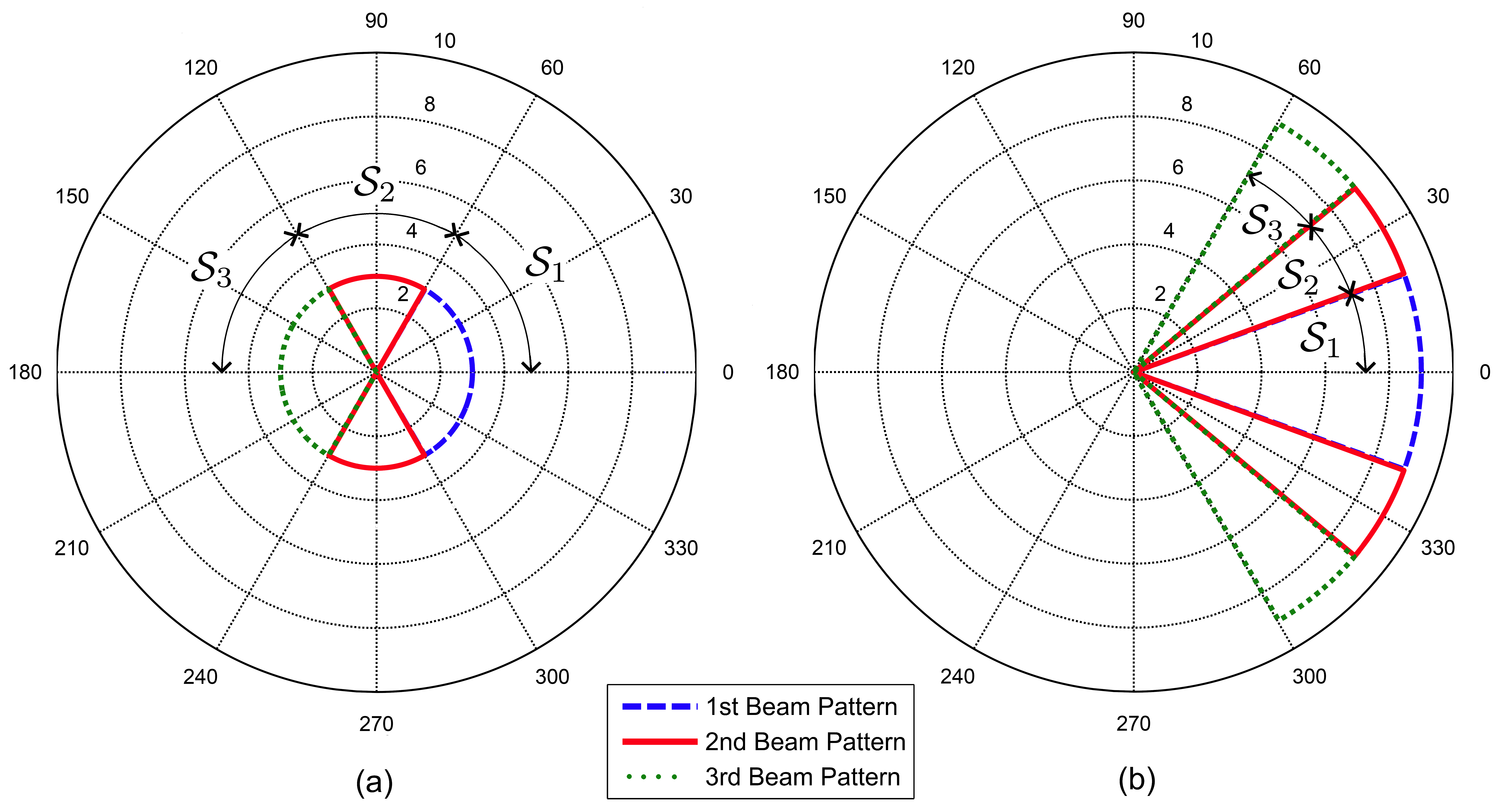}
\caption{Illustration of the beam patterns adopted in the first (a) and second (b) stages of the channel estimation algorithm of \cite{rheath} when $K = 3$. The three sub-ranges in the first stage are, $[0, \pi/3)$, $[\pi/3, 2\pi/3)$ and $[2\pi/3, \pi)$, respectively. By assuming that the possible AOAs/AODs are reduced to the sub-range $[0, \pi/3)$ in the first stage, this sub-range is further divided into $[0, \pi/9)$, $[\pi/9, 2\pi/9)$ and $[2\pi/9, \pi/3)$, respectively, in the second stage.}
\label{rheath_model}
\end{figure}

As one of the seminal works on multi-stage codebook approaches, \cite{rheath} developed a ``divide and conquer" search type algorithm to estimate sparse mmWave channels. As shown in Fig. \ref{rheath_model}, in each stage of this algorithm, the possible ranges of angles of departure (AODs) and angles of arrival (AOAs) are both divided into $K<<N$ non-overlapped angular sub-ranges. Correspondingly, $K$ non-overlapped beam patterns are designed at both the transmitter and receiver such that each transmit (receive) beam pattern exactly covers one AOD (AOA) sub-range. The channel estimation carried out in each stage consists of $K^2$ time slots. In each time slot, the pilot signal is transmitted using one of the $K$ beam patterns at the transmitter, and then collected by one of the $K$ beam patterns at the receiver. The corresponding channel output for each pair of transmit-receive beam patterns can then be obtained. These $K^2$ time slots span all the combinations of transmit-receive beam patterns. By comparing the magnitudes of the corresponding $K^2$ channel outputs, the transmit/receive sub-ranges that the AOD/AOA  most likely  belong to are determined. The receiver can then feed back the AOD information for use at the transmitter. Afterwards, the algorithm will limit the estimation to the angular sub-range identified at each link end in the previous stage and further divide it into $K$ sub-ranges for the channel estimation in the next stage. {\color{\col} An example of multi-stage angular refinement, using our proposed approach, can be seen in Fig. \ref{high_level}}. This process continues until the smallest beam width resolution is reached. It is shown in \cite{rheath} that the algorithm requires estimation time proportional to $K^2\lceil \text{log}_K (N)\rceil$ per path. Despite the significant improvement when compared to the exhaustive search approach, such a channel estimation algorithm still might not be quick enough to track the fast channel variations, especially for mmWave mobile channels with rapidly changing parameters. Furthermore, at high SNR it may not be necessary to perform so many measurements, which would result in an unnecessary time delay. {\color{\col}Adaptive training approaches are also investigated in \cite{xia2008multi,xiao2015iterative,xiao2014iterative,xia2008practical}. While these approaches are shown to significantly improve system performance as the number of measurement iterations is increased, there is generally no adaptation of the number of measurements with respect to the associated probability of success or channel conditions.}

Motivated by this, in this paper we develop a fast mmWave MIMO channel estimation framework by designing a set of novel overlapped beam patterns that can significantly reduce the number of channel estimation measurements. In order to improve estimation accuracy, we then introduce a novel rate-adaptive channel estimation approach, where the average number of channel measurements is adapted to channel conditions. The main contributions of this paper are summarized as follows:

\begin{itemize}
\item{
Relying on novel overlapped beam pattern design, we first present a fast channel estimation (FCE) algorithm for mmWave systems. In this algorithm, we develop a maximum likelihood (ML) detector to optimally extract the channel AOD/AOA information from the measurements. We also design a linear minimum mean squared error (LMMSE) channel estimator to estimate the channel coefficients by optimally combining the selected measurements in all stages.
}
\item{We then develop a rate-adaptive channel estimation (RACE) algorithm, in which additional measurements are permitted when the current measurements are found to have an inadequate probability of success. In this way, the number of measurements can adapt to the channel conditions and the channel estimation accuracy can be significantly improved with minimal measurements.
}
\item{
We analyze the probability of channel estimation error for the FCE algorithm. In particular, we derive a closed-form approximation, lower bound and upper bound for the PEE. Based on the Shannon-Hartley theorem, we also provide some theoretical analysis for the minimum energy required to estimate the channel using the RACE algorithm.
}
\item{
Finally, we compare the performance of the proposed algorithms to that of the algorithm in \cite{rheath} with non-overlapped beam patterns.
Simulation results show that both of the proposed algorithms can significantly reduce the number of channel measurements compared to \cite{rheath}. We show that the FCE algorithm achieves a guaranteed reduction of channel measurements, at the expense of estimation accuracy. On the other hand, the RACE algorithm can achieve the same average reduction of channel measurements as the FCE algorithm, but using up to 6dB less signal energy compared to the algorithm in \cite{rheath}.
}
\end{itemize}

\textit{Notation} : $\boldsymbol{A}$ is a matrix, $\boldsymbol{a}$ is a vector, ${a}$ is a scalar, and $\mathcal{A}$ is a set. $||\boldsymbol{A}||_2$ is the 2-norm of $\boldsymbol{A}$, $\text{det}(\boldsymbol{A})$ is the determinant of $\boldsymbol{A}$ and $|\mathcal{A}|$ represents the cardinality of $\mathcal{A}$. $\boldsymbol{A}^T$, $\boldsymbol{A}^H$ and $\boldsymbol{A}^*$ are the transpose, conjugate transpose and conjugate of $\boldsymbol{A}$, respectively. For a square matrix $\boldsymbol{A}$, $\boldsymbol{A}^{-1}$ represents its inverse. We use $\text{diag}({\boldsymbol{a}})$ to denote a diagonal matrix with entries of $\boldsymbol{a}$ on its diagonal. $\boldsymbol{I}_N$ is the $N\times N$ identity matrix, $\boldsymbol{1}_N$ is an $N\times 1$ all $1$ column vector and $\lceil \cdot \rceil$ denotes the ceiling function. $\mathcal{C}\mathcal{N}(\boldsymbol{m},\boldsymbol{R})$ is a complex Gaussian random vector with mean $\boldsymbol{m}$ and covariance matrix $\boldsymbol{R}$. $\text{E}[\boldsymbol{a}]$ and $\text{Cov}[\boldsymbol{a}]$ denote the expected value and covariance of ${\boldsymbol{a}}$, respectively. $\boldsymbol{A} \otimes \boldsymbol{B}$ denotes the Kronecker product of $\boldsymbol{A}$ and $\boldsymbol{B}$ whereas $\boldsymbol{A} \odot \boldsymbol{B}$ denotes the row-wise Kronecker product of $\boldsymbol{A}$ and $\boldsymbol{B}$.

\section{System Model}

Consider a mmWave MIMO system composed of $N_t$ transmit antennas and $N_r$ receive antennas. We consider that both the transmitter and receiver are equipped with a limited number of radio frequency (RF) chains. Following \cite{rheath}, we further assume that these RF chains, at one end, can only be combined to form a single beam pattern, indicating that only one pilot signal can be transmitted and received at one time. {\color{\col}In this paper, for simplicity, we consider the unconstrained beamforming vectors by ignoring some practical constraints imposed by hardware such as constant amplitude and quantized phase shifters. However, in practice, our unconstrained beamforming vectors could be realized by using a network of constrained beamformers with quantized phase shifters and constant amplitude as the hybrid-beamforming approach adopted in \cite{rheath} and depicted by Figure 2 therein.} To estimate the channel matrix, the transmitter sends a pilot signal $x$, with unit energy ($||x||_2=1$), to the receiver. Denote by $\boldsymbol{f}$ and $\boldsymbol{w}$  $(||\boldsymbol{f}||_2 = ||\boldsymbol{w}||_2 = 1)$, respectively, the $N_t \times 1$ beamforming vector at the transmitter and $N_r \times 1$ beamforming vector at the receiver. The corresponding channel output can be represented as
\begin{equation}
\label{y}
y = \sqrt{P} \boldsymbol{w}^H \boldsymbol{H} \boldsymbol{f} x + \boldsymbol{w}^{H} \boldsymbol{q},
\end{equation}
\noindent
where $\boldsymbol{H}$ denotes the $N_r\times N_t$ MIMO channel matrix, $P$ is the transmit power and $\boldsymbol{q}$ is an $N_r \times 1$ complex additive white Gaussian noise (AWGN) vector following distribution $\mathcal{C}\mathcal{N}(0, N_0 \boldsymbol{I}_{N_r})$.

%
%

In this paper we follow \cite{Sayeed_max} and adopt a two-dimensional (2D) sparse geometric-based channel model. Specifically, we consider an $L$-path channel between the transceiver, with the $l$th path having steering AOD, $\phi^t_l$, and AOA, $\phi^r_l$ where $l=1,...,L$. Then the corresponding channel matrix can be expressed in terms of the physical propagation path parameters as
%
\begin{equation}
\label{H_sum}
\boldsymbol{H} = \sqrt{N_tN_r}\sum\limits_{l=1}^L\alpha_l  \boldsymbol{a}_{r}(\phi_l^r) \boldsymbol{a}_{t}^{H}(\phi_l^t)
\end{equation}
\noindent
where $\alpha_l$ is the fading coefficient of the $l$th propagation path, and $\boldsymbol{a}_t(\phi_l^t)$ and $\boldsymbol{a}_r(\phi_l^r)$ respectively denote the transmit and receive spatial signatures of the $l$th path. To simplify the analysis, we assume that  the transmitter and receiver have the same number of antennas (i.e., $N_t = N_r = N$). However, it is worth pointing out that the developed schemes can be easily extended to a general asymmetric system. If uniform linear antenna arrays (ULA) are employed at both the transmitter and receiver, we can define $\boldsymbol{a}_t(\phi_l^t)= \boldsymbol{u}(\phi_l^t)$ and $\boldsymbol{a}_r(\phi_l^r)= \boldsymbol{u}(\phi_l^r)$, respectively, where
%
\begin{equation}
\label{u_n}
\boldsymbol{u}(\epsilon) \triangleq \frac{1}{\sqrt{N}} [1,e^{j 2 \pi \epsilon},\cdots,e^{j2 \pi (N-1)\epsilon}]^T.
\end{equation}
\noindent
Here, the steering angle, $\phi_l^t$, is related to the physical angle $\theta_l^t \in [0,\pi)$ by $\phi_l^t=\frac{d\text{ sin}(\theta_l^t)}{\lambda}$ with $\lambda$ denoting the signal wavelength\footnote{Note that the use of ULA results in no distinguishable difference between AODs $\theta_l^t$ and $-\theta_l^t$ or between AOAs $\theta_l^r$ and $-\theta_l^r$. Hence, only AODs and AOAs in the range $[0,\pi)$ need to be considered.}. A similar expression can be written for $\phi_l^r$ at the receiver. With half-wavelength spacing, the distance between antenna elements becomes $d=\lambda/2$.

From (\ref{H_sum}), we can see that the overall channel state information of each path includes only three parameters, i.e., the AOD $\phi_l^r$, the AOA $\phi_l^r$, and the fading coefficient $\alpha_l$. We assume that the fading coefficient of each path follows a complex Gaussian distribution with zero mean and variance $P_R$ and that both $\phi_l^t$ and $\phi_l^r$ can only take some discrete values from the set $\mathcal{U}_N=\{0,\frac{\pi}{N},\cdots,\frac{\pi(N-1)}{N} \}$. {\color{\col}Here, for the sake of ensuing mathematical problem formulation, we only consider the discrete AOA and AOD. It is noteworthy that they can be continuous in practice. However, the extension to the case with continuous AOD/AOA may require the consideration of other more practical issues such as the number of RF chains to realize the beam patterns and the hardware constraints (e.g., quantized phase shifters) imposed on the RF beamforming vectors, which may constitute a new paper. We thus have left this extension as our future work.}

We aim to find an efficient way to estimate the three parameters for each path. The key challenge here is how to design a sequence of $\boldsymbol{f}'s$ and $\boldsymbol{w}'s$ in such a way that the channel parameters can be quickly and accurately estimated. We consider $M$ pairs of beam patterns that are designed to span all possible transmit-receive combinations. Denote by $\boldsymbol{f}_m$ and $\boldsymbol{w}_m$, respectively, the transmit and receive beamforming vectors adopted in the $m$th channel measurement time slot such that $||\boldsymbol{f}_m||_2=||\boldsymbol{w}_m||_2 = 1,\;\forall\;m=1,\cdots,M$. Similarly to \cite{rheath}, we assume the same pilot symbol $x$ is transmitted during the $M$ time slots. Then, after $M$ time slots, we can obtain a sequence of $M$ measurements represented as
\begin{align}
\label{y_h_v}
\boldsymbol{y} =& \sqrt{P} x \boldsymbol{h}_v +\boldsymbol{n},
\end{align}
where $\boldsymbol{h}_v$ describes the channel input-output relationship for a given set of transmit and receive beamforming vectors defined by
\begin{align}
\label{h_v}
    \boldsymbol{h}_v  =
    \left[\begin{array}{ccc}
   	\boldsymbol{w}_1^H \boldsymbol{H} \boldsymbol{f}_1  \\
   	\boldsymbol{w}_2^H \boldsymbol{H} \boldsymbol{f}_2  \\
   	\vdots												\\ 	
   	\boldsymbol{w}_M^H \boldsymbol{H} \boldsymbol{f}_M
    \end{array}\right]
\end{align}
and
\begin{align}
\label{n_q}
    \boldsymbol{n}  =
    \left[\begin{array}{ccc}
   	\boldsymbol{w}^H_{1} \boldsymbol{q}_1   \\
   	\boldsymbol{w}^H_{2} \boldsymbol{q}_2   \\
   	\vdots									\\ 	
   	\boldsymbol{w}^H_{M} \boldsymbol{q}_M
    \end{array}\right]
\end{align}
\noindent
is an $M\times 1$ vector of the corresponding noise terms. Note that  since $||\boldsymbol{w}_m||_2 = 1$, $\forall$ $m$, the vector $\boldsymbol{n}$ follows the same distribution as that of ${\boldsymbol{q}_m}$, i.e., $\boldsymbol{n} \sim \mathcal{C}\mathcal{N}(0, N_0I_M)$.

Motivated by the geometric sparsity of the mmWave channel, in the following two sections, we propose to use a set of overlapped beam patterns that are able to estimate the AOD/AOA information very quickly. We then extend the algorithm to use a rate-adaptive estimation approach. The adaptive nature of the algorithm permits additional measurements to be performed under poor channel conditions. This allows the fast channel estimation to be carried out with significant accuracy and energy efficiency.

\section{Fast Channel Estimation with Overlapped Beam Patterns}

In this section, we develop a fast channel estimation framework for mmWave systems using overlapped beam patterns, {\color{\col} as illustrated in Fig. \ref{high_level}}. Specifically, we design a set of beam patterns that are adopted in different measurement time intervals and are overlapped with one another in the angular domain. A maximum likelihood based estimation algorithm is then proposed to accurately retrieve the channel state information from the set of measurements. The proposed channel estimation algorithm also works in a similar multi-stage manner as that in \cite{rheath} where each stage reduces the possible sub-ranges in which the AOD/AOA are expected to be found.

\subsection{An Example of Overlapped Beam Pattern Design} \label{sssec:example_beam_patterns}
\begin{figure}[!t]
\centering
\includegraphics[width=3.5in]{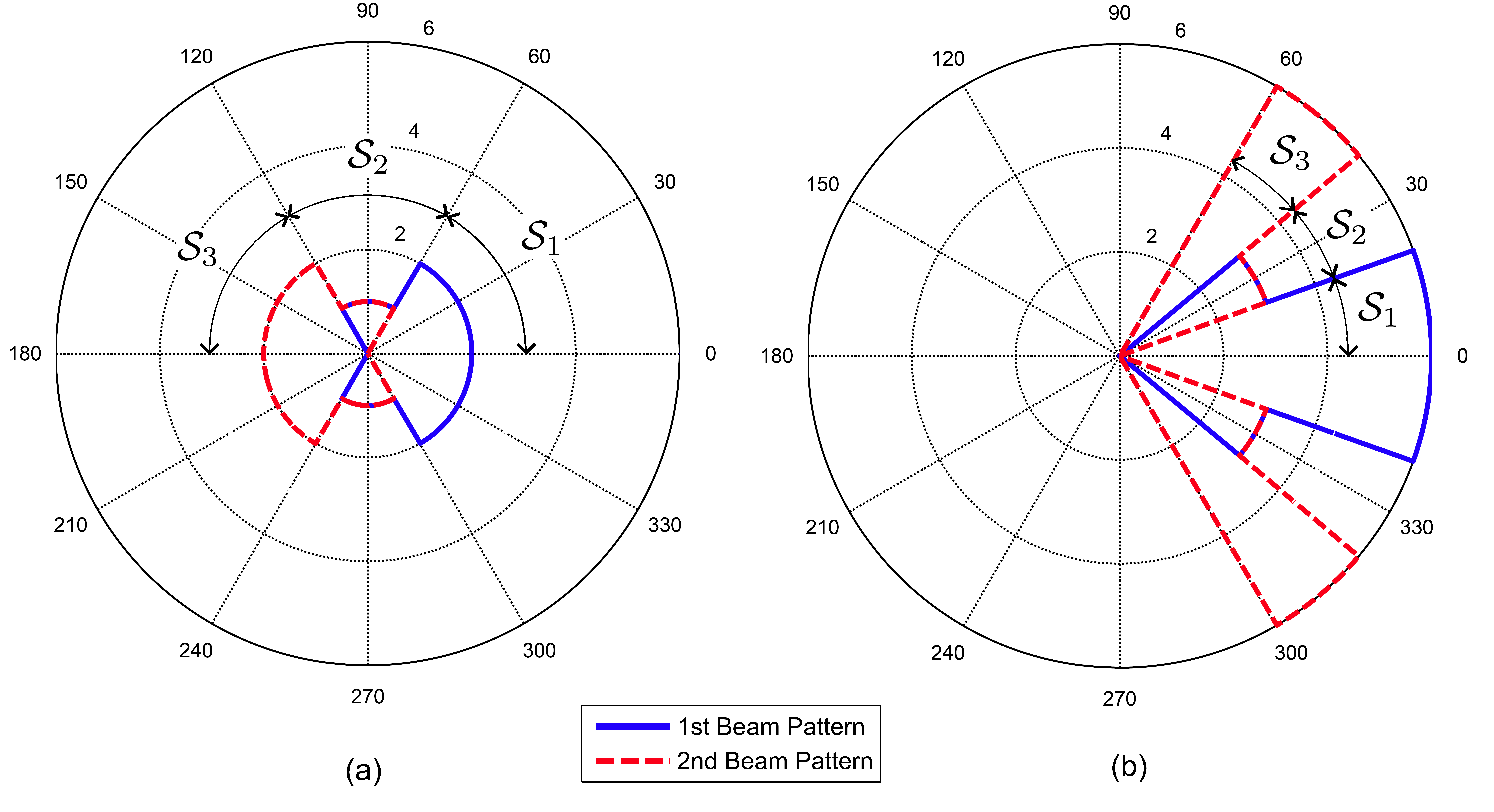}
\caption{Illustration of the overlapped beam patterns adopted in the first (a) and second (b) stages of the proposed algorithm when $K = 3$. By assuming that the possible AOAs/AODs are reduced to the sub-range $[0, \pi/3)$ in the first stage, this sub-range is further divided into $[0, \pi/9)$, $[\pi/9, 2\pi/9)$ and $[2\pi/9, \pi/3)$, respectively, in the second stage.}
\label{US_K3}
\end{figure}

\begin{figure*}[!t]
\centering
\includegraphics[width=7.0in,trim={3.0cm 2cm 3cm 3.5cm},clip]{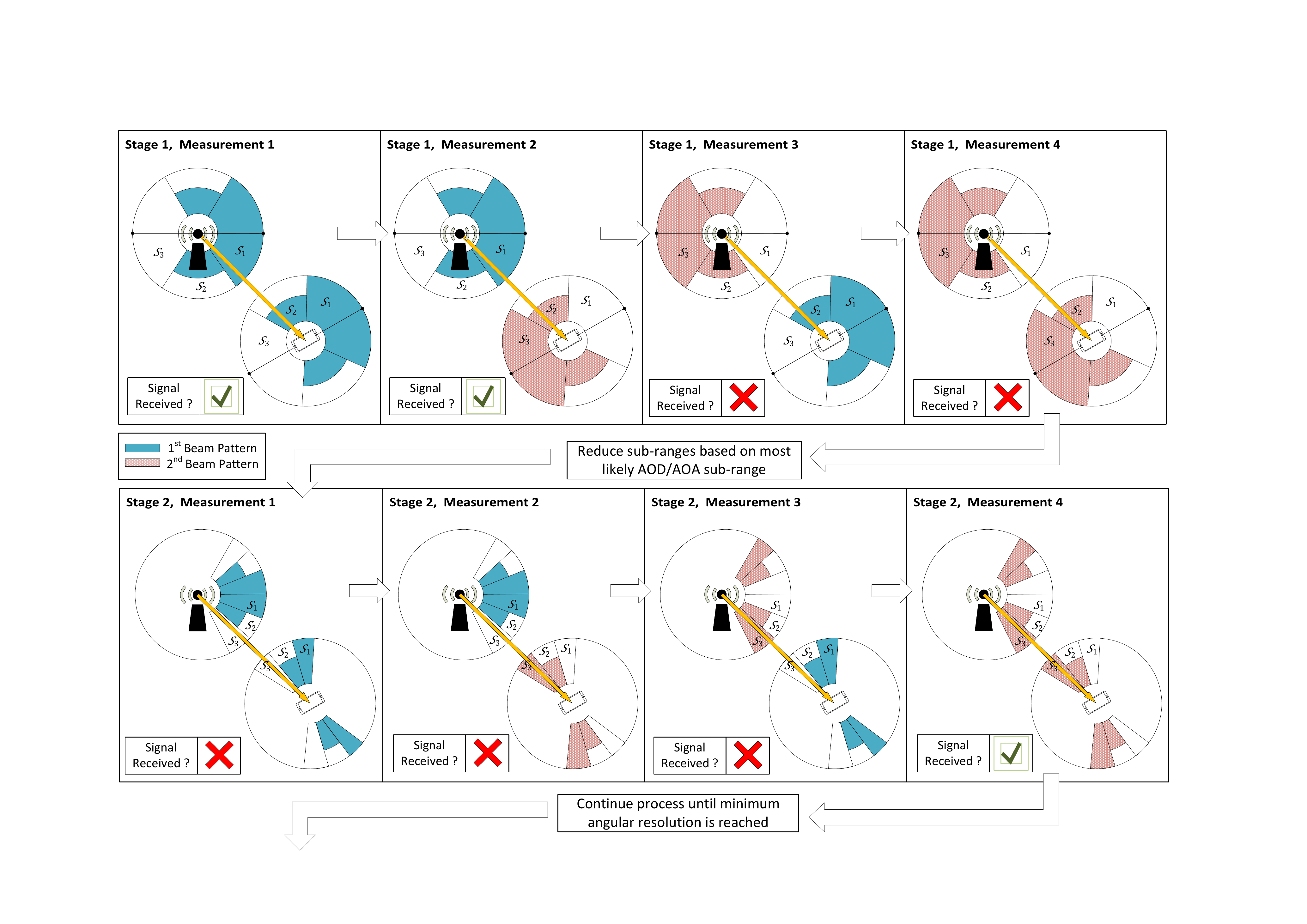}
\caption{\color{\col}Illustration of multi-stage channel estimation using the example overlapped beam patterns. Here we use an arrow to represent the propagation path to be estimated. As can be seen, in the first stage, it is expected that a signal will only be received in the first and second measurements (due to the propagation path angle). As the transmit beam patterns used are predetermined and thus known to the receiver, it can be logically deduced that the AOD at the transmitter can only belong to the 1st sub-range and that the AOA at the receiver should belong to the 2nd sub-range. It is noteworthy that this series of measurements (i.e., [\cmark,\cmark,\xmark,\xmark]) corresponds to the 2nd column of the generator matrix in (\ref{G_K_3_2}). The receiver then feeds back the AOD sub-range to the transmitter for the channel estimation in next stage. Both the transceiver and receiver then divide their estimated sub-range into narrower sub-ranges and carry out further overlapped sub-range measurements. As can be seen, in the second stage, it is expected that a signal will only be received in the final measurement. Logically, this means that the AOD at the transmitter is within the 3rd sub-range and the AOA at the receiver is on the 3rd sub-range. This series of measurements (i.e., [\xmark, \xmark, \xmark, \cmark]) also corresponds to the 9th column of the generator matrix in (\ref{G_K_3_2}).}
\label{high_level}
\end{figure*}

We will first explain the design principle of overlapped beam patterns using a simple example. Following Fig. \ref{rheath_model}, we divide the  AOD/AOA angular spaces into $K = 3$  sub-ranges in the first stage, denoted by $\mathcal{S}_{1} =\{\epsilon \in \mathcal{U}_N|  0 \leq \epsilon <\pi/3 \}$, $\mathcal{S}_{2} = \{\epsilon \in \mathcal{U}_N|  \pi/3 \leq \epsilon < 2\pi/3 \}$ and $\mathcal{S}_{3} = \{\epsilon \in \mathcal{U}_N|  2\pi/3 \leq \epsilon <\pi \}$, respectively. However, instead of using 3 beam patterns to cover them at each transceiver end as in Fig. \ref{rheath_model}(a), we propose to use only 2 overlapped beam patterns to achieve this. Fig. \ref{US_K3}(a) illustrates our designed beam patterns in the first stage. We can see that the first and second beam patterns cover $\mathcal{S}_1$, $\mathcal{S}_2$ and $\mathcal{S}_2$, $\mathcal{S}_3$, respectively, and are overlapped in the whole range of $\mathcal{S}_2$. Intuitively, if a path is observed in two measurements using adjacent beam patterns, the AOD or AOA must belong to the overlapped sub-range of these two beam patterns. It is also seen that each beam pattern can have different amplitudes in different sub-ranges. We represent the amplitudes of each beam pattern in different sub-ranges by a vector. For beam pattern 1 and 2 in Fig. \ref{US_K3}(a), these vectors are respectively defined as
%
\begin{equation}
\label{B_generic_1}
    \boldsymbol{b}_1 =
    \left[\begin{array}{ccc}
    b_{1,1}, &b_{1,2}, &b_{1,3}\\
    \end{array}\right],\;\\
    \boldsymbol{b}_2 =
    \left[\begin{array}{ccc}
    b_{2,1}, &b_{2,2}, &b_{2,3}
    \end{array}\right]
\end{equation}
\noindent
where $\boldsymbol{b}_i$ corresponds to the $i$th beam pattern with $b_{i,k}$ denoting the amplitude of the $i$th beam pattern in sub-range $S_{k}  , \; \forall \; k=1,...,K$. By using $M=4$ measurement time slots, we can then span all beam pattern combinations between the transceiver. We denote the sequential set of beam patterns respectively adopted at the transmitter and receiver by
%
\begin{align}
\label{B_generic_2}
    \boldsymbol{B}_T^{(M)}  =
    \left[\begin{array}{ccc}
   	\boldsymbol{b}_1 \\
   	\boldsymbol{b}_1 \\
   	\boldsymbol{b}_2 \\
   	\boldsymbol{b}_2
    \end{array}\right]=
    \left[\begin{array}{ccc}
   	b_{1,1}, &b_{1,2}, &b_{1,3} \\
   	b_{1,1}, &b_{1,2}, &b_{1,3} \\
   	b_{2,1}, &b_{2,2}, &b_{2,3} \\
   	b_{2,1}, &b_{2,2}, &b_{2,3}
    \end{array}\right], \\
    \label{B_generic_22}
    \boldsymbol{B}_R^{(M)}  =
    \left[\begin{array}{ccc}
   	\boldsymbol{b}_1 \\
   	\boldsymbol{b}_2 \\
   	\boldsymbol{b}_1 \\
   	\boldsymbol{b}_2 \\
    \end{array}\right] =
    \left[\begin{array}{ccc}
   	b_{1,1}, &b_{1,2}, &b_{1,3} \\
   	b_{2,1}, &b_{2,2}, &b_{2,3} \\
   	b_{1,1}, &b_{1,2}, &b_{1,3} \\
   	b_{2,1}, &b_{2,2}, &b_{2,3}
    \end{array}\right].
\end{align}
\noindent
We refer to these as the beam pattern design matrices, with their $m$th row denoting the beam pattern adopted in the $m$th measurement time slot, where $m=1,...,M$. The efficient design of these beam patterns can lead to many solutions. However, one desirable property is that the same quantity of signal energy is transmitted/received via each sub-range over all measurements, i.e., the energy of each column of (\ref{B_generic_2})-(\ref{B_generic_22}) should have the same Euclidean norm. This provides the same accuracy for each possible sub-range combination. Another desirable property is that the transmit/receive beamforming gains of all measurement patterns are equal, i.e., the energy of each rows of (\ref{B_generic_2})-(\ref{B_generic_22}) should be the same.

One possible way to make the beam pattern sub-range amplitudes in (\ref{B_generic_1}) follow the aforementioned properties, is to normalize $\boldsymbol{B}^{(M)}_T$ and $\boldsymbol{B}^{(M)}_R$ to have unit energy in each row and have equal energy in each column. For the beam patterns in Fig. \ref{US_K3}, we have $b_{1,3}=b_{2,1}=0$, as beam patterns $1$ and $2$ do not cover, respectively, the third and first sub-ranges. Due to the symmetry between the two beam patterns, we further have $b_{1,1} = b_{2,3} = \beta_1$ and $b_{1,2} = b_{2,2} = \beta_2$. This leads to $\beta_1 = \frac{{1}}{\sqrt{3}}$ and $\beta_2 = \frac{\sqrt{2}}{\sqrt{3}}$, and the matrices in (\ref{B_generic_2})-(\ref{B_generic_22}) become
\begin{align}
\label{B_full}
    \boldsymbol{B}_T^{(M)}  =
    \left[\begin{array}{ccc}
    \frac{\sqrt{2}}{\sqrt{3}}&\frac{{1}}{\sqrt{3}}&0\\
    \frac{\sqrt{2}}{\sqrt{3}}&\frac{{1}}{\sqrt{3}}&0\\
    0&\frac{{1}}{\sqrt{3}}&\frac{\sqrt{2}}{\sqrt{3}}\\
    0&\frac{{1}}{\sqrt{3}}&\frac{\sqrt{2}}{\sqrt{3}}
    \end{array}\right],\;
    \boldsymbol{B}_R^{(M)}  =
    \left[\begin{array}{ccc}
    \frac{\sqrt{2}}{\sqrt{3}}&\frac{{1}}{\sqrt{3}}&0\\
    0&\frac{{1}}{\sqrt{3}}&\frac{\sqrt{2}}{\sqrt{3}}\\
    \frac{\sqrt{2}}{\sqrt{3}}&\frac{{1}}{\sqrt{3}}&0\\
    0&\frac{{1}}{\sqrt{3}}&\frac{\sqrt{2}}{\sqrt{3}}
    \end{array}\right]. \nonumber \\
\end{align}
In order to observe the resultant amplitude gains (i.e., the transceiver gain) over each of the $K^2=9$ sub-range combinations, we introduce another matrix referred to as the generator matrix. We define this as the row-wise Kronecker product between the transmit beam pattern design matrix and the receive beam pattern design matrix such that
%

{\color{\col}
\begin{align}
\label{G_K_3}
    \boldsymbol{G}^{(M)}&= \boldsymbol{B}_T^{(M)} \odot  \; \boldsymbol{B}_R^{(M)}
    =\left[\begin{array}{ccc}
    \boldsymbol{b}_1 \otimes \boldsymbol{b}_1\\
    \boldsymbol{b}_1 \otimes \boldsymbol{b}_2\\
    \boldsymbol{b}_2 \otimes \boldsymbol{b}_1\\
    \boldsymbol{b}_2 \otimes \boldsymbol{b}_2\\
    \end{array}\right]  \\ \nonumber \\
    &=\frac{1}{3}\left[\begin{array}{ccccccccc}
    2&\sqrt{2}&0&\sqrt{2}&{1}&0&0&0&0\\
    0&\sqrt{2}&2&0&1&\sqrt{2}&0&0&0\\
    0&0&0&\sqrt{2}&1&0&2&\sqrt{2}&0\\
    0&0&0&0&1&\sqrt{2}&0&\sqrt{2}&2
    \end{array}\right] \label{G_K_3_2}, \nonumber \\
\end{align}
\noindent
recalling} that $\odot$ and $\otimes$ represent the row-wise Kronecker and Kronecker product operations, respectively. By denoting ${b}^{(m,k^t)}_{T}$ and ${b}^{(m,k^r)}_{R}$ as the entry on the $k^t$th and $k^r$th column on the $m$th row in $\boldsymbol{B}_T^{(M)}$ and $\boldsymbol{B}_R^{(M)}$, respectively, we can express the $d$th column of $\boldsymbol{G}^{(M)}$ as
\begin{align}
\label{G_col}
    \boldsymbol{G}^{(M)}_d=   \left[\begin{array}{ccc}
   	{b}^{(1,k^t)}_{T} {b}^{(1,k^r)}_{R} \\
   	\vdots										\\ 	
   	{b}^{(M,k^t)}_{T} {b}^{(M,k^r)}_{R} \\
    \end{array}\right]
\end{align}
where the relationship $d=K(k^t-1)+k^r$ is a result of the Kronecker product operation.

The columns of generator matrix describe the $M=4$ received measurement gains over each of the $K^2=9$ sub-range combinations. For example, if a path were present between the transmit sub-range $k^t=3$ and the receive sub-range $k^r=1$, the measurement vector $\boldsymbol{y}$ would be expected to be a scalar multiple of column $d=7$ of $\boldsymbol{G}^{(M)}$, expressed as $\boldsymbol{G}^{(M)}_7$. It is important to note that the sets of beam patterns used in this paper are not unique. We later show that their performance, in terms of PEE, depends only on the Euclidean distance between each column of $\boldsymbol{G}^{(M)}$, which directly determines the probability that one column corresponding to a certain sub-range is to be mistaken for another. In the example set shown in (\ref{G_K_3_2}), each column has the same equal minimum Euclidean distance when compared with all other columns, although some have more spatial neighbours at this minimum distance than others.

\subsection{Beamforming Vector Design}
To generate the beam patterns illustrated in Fig. \ref{US_K3}(a) and described in (\ref{B_full}), the transmit and receive beamforming vectors should be designed as follows. Denote by $\boldsymbol{f}_m$ and $\boldsymbol{w}_m$, respectively, the transmit beamforming vector and receive beamforming vector corresponding to the $m$th pair of beam patterns in $\boldsymbol{B}^{(M)}_T$ and $\boldsymbol{B}^{(M)}_R$. We then design the product of the transmit array response and transmit beamforming vector to have
\begin{equation}
\label{f_example}
\boldsymbol{u}^H(\epsilon) \boldsymbol{f}_m  =  C {b}^{(m,k)}_{T}, \text{ if $ \exists \; k \in  \{ 1, 2,..., K \},\epsilon \in  \mathcal{S}_{k}  $},
\end{equation}
\noindent
and the product of the receiver array response and receive beamforming vector to have
\begin{equation}
\label{w_example}
\boldsymbol{u}^H(\epsilon) \boldsymbol{w}_m =  C {b}^{(m,k)}_{R}, \text{ if $ \exists \; k \in \{ 1, 2,..., K \},\epsilon \in \mathcal{S}_{k}  $},
\end{equation}

\noindent
where $\boldsymbol{u}(\epsilon)$ has been defined in (\ref{u_n}) and $C$ is a scalar constant that ensures $||\boldsymbol{f}_m||_2=||\boldsymbol{w}_m||_2=1$. Physically, $C$ corresponds to the average directivity gain of each beam pattern and is the same for all $m=1,\cdots,M$ due to the normalization of the rows in (\ref{B_full}). Eqs.  (\ref{f_example}) and (\ref{w_example}) can be expressed in a matrix form as
\begin{equation}
\label{matrix_eq_1}
\boldsymbol{U}^H \boldsymbol{f}_m = \left[\begin{array}{ccc}
    C{b}^{(m,1)}_{T} \boldsymbol{1}_{|S_{1}|} \\
    \vdots \\
    C{b}^{(m,K)}_{T} \boldsymbol{1}_{|S_{K}|}
    \end{array}\right]
     \triangleq \boldsymbol{z}_T^{(m)}
\end{equation}
and
\begin{equation}
\label{matrix_eq_2}
\boldsymbol{U}^H \boldsymbol{w}_m = \left[\begin{array}{ccc}
    C{b}^{(m,1)}_{R} \boldsymbol{1}_{|S_{1}|} \\
    \vdots \\
    C{b}^{(m,K)}_{R} \boldsymbol{1}_{|S_{K}|}
    \end{array}\right]
     \triangleq \boldsymbol{z}_R^{(m)}
\end{equation}
\noindent
where $\left| \mathcal{S} \right|$ denotes the cardinality of set $\mathcal{S}$ and $\boldsymbol{U} = \big[ \boldsymbol{u}(0),\boldsymbol{u} \Big( \frac{\pi}{N} \Big),\cdots,\boldsymbol{u} \Big( \frac{\pi(N-1)}{N} \Big) \big]$ is a matrix whose columns describe the antenna array response at each angle. Therefore $\boldsymbol{f}_m$ and $\boldsymbol{w}_m$ can be designed as
\begin{align}
\label{f_design}
\boldsymbol{f}_m = (\boldsymbol{U}\boldsymbol{U}^H)^{-1}\boldsymbol{U} \boldsymbol{z}_T^{(m)}
\end{align}
and
\begin{align}
\label{w_design}
\boldsymbol{w}_m = (\boldsymbol{U}\boldsymbol{U}^H)^{-1}\boldsymbol{U} \boldsymbol{z}_R^{(m)}
\end{align}
where $(\boldsymbol{U}\boldsymbol{U}^H)^{-1}\boldsymbol{U}$ is the pseudo inverse of $\boldsymbol{U}$.

\subsection{Channel Measurements}
We now perform channel estimation in the first stage using the previously designed transmit and receive beamforming vectors $\{\boldsymbol{f}_m\}$ and $\{\boldsymbol{w}_m\}$. In each time slot, the beamforming vectors $\boldsymbol{f}_m$ and $\boldsymbol{w}_m$ are adopted to transmit/receive the pilot signal $x$. If we substitute the channel in (\ref{H_sum}) into (\ref{h_v}), we get
\begin{align}
\label{h_v_2}
    \boldsymbol{h}_v^{(M)}  = x N \sum\limits_{l=1}^L\alpha_l
    \left[\begin{array}{ccc}
   	(\boldsymbol{a}^H_{r}(\phi_l^r) \boldsymbol{w}_1 )^H \boldsymbol{a}_{t}^{H}(\phi_l^t) \boldsymbol{f}_1  \\
   	\vdots												\\ 	
   	(\boldsymbol{a}^H_{r}(\phi_l^r) \boldsymbol{w}_M )^H \boldsymbol{a}_{t}^{H}(\phi_l^t) \boldsymbol{f}_M
    \end{array}\right].
\end{align}
\noindent
Without loss of generality, let us consider the case with AOD, $\phi_l^t \in \mathcal{S}_{k_l^t}$ and AOA, $\phi_l^r \in \mathcal{S}_{k_l^r}$, where $k_l^t$ and $k_l^r$ are respectively, the transmit and receive sub-range indices of the $l$th propagation path. By recalling (\ref{f_example})-(\ref{w_example}) we write
\begin{eqnarray}
\label{h_v_simplify}
\boldsymbol{u}^H(\phi_l^t) \boldsymbol{f}_m = C {b}^{(m,k_l^t)}_{T} \text{ and } \boldsymbol{u}^H(\phi_l^r) \boldsymbol{w}_m = C {b}^{(m,k_l^r)}_{R}
\end{eqnarray}
\noindent
which leads to
\begin{align}
\label{h_v_3}
    \boldsymbol{h}_v^{(M)}  = x N C^2 \sum\limits_{l=1}^L\alpha_l
    \left[\begin{array}{ccc}
   	{b}^{(1,k_l^t)}_{T} {b}^{(1,k_l^r)}_{R} \\
   	\vdots										\\ 	
   	{b}^{(M,k_l^t)}_{T} {b}^{(M,k_l^r)}_{R} \\
    \end{array}\right].
\end{align}
\noindent
We can see from (\ref{G_col}) that the vector term in (\ref{h_v_3}) is the weighted sum of the columns of $\boldsymbol{G}^{(M)}$ i.e., the weighted sum of columns $d_l=K(k_l^t-1)+k_l^r, \; \forall \; l=1,...,L$, in $\boldsymbol{G}^{(M)}$. Therefore we can express $\boldsymbol{h}_v$ by the generator matrix as
\begin{align}
\label{h_v_r}
    \boldsymbol{h}_v^{(M)}  = x N C^2
    \boldsymbol{G}^{(M)} \boldsymbol{v}^T
\end{align}
\noindent
where $\boldsymbol{v}$ is a $1 \times K^2$ sparse row vector that describes the channel gain at each of the $K^2$ sub-range combinations by
\begin{align}
\label{v}
     v_{d_l} = \alpha_{l} ,\; \forall \; l=1,...,L
\end{align}
\noindent
and zero otherwise. For example with $K=3$, a single path (i.e., $L=1$) with coefficient $\alpha_1$, exists on the first transmit sub-range (i.e., $k_1^t=1$) and second receive sub-range (i.e., $k_1^r=2$) leads to $d_1=K(k_1^t-1)+k_1^r=2$ and
\begin{align}
\label{v_eg}
     \boldsymbol{v} =  \{0,\alpha_{1},0,0,0,0,0,0,0\}.
\end{align}
\noindent
Finally by using (\ref{h_v_r}), we can re-write the received channel output vector defined in (\ref{y}), after $M$ measurements, as
\begin{align}
\label{y_g}
\boldsymbol{y}^{(M)} =& \sqrt{P}N C^2 x \boldsymbol{G}^{(M)} \boldsymbol{v}^T +\boldsymbol{n}^{(M)}.
\end{align}

\subsection{Maximum Likelihood Detection of AOD/AOA Information}
We now require an efficient means of detecting $\boldsymbol{v}$ given that a generator matrix $\boldsymbol{G}^{(M)}$ has been used to obtain the channel outputs in (\ref{y_g}). Due to its optimal detection properties, this subsection elaborates how to implement a maximum likelihood detection \cite{proakis} method to extract the AOD/AOA information from the received measurements. We begin by considering the distribution of $\boldsymbol{y}^{(M)}$. From (\ref{y_g}), this can be expressed as
\begin{align}
\label{y_dist}
    \boldsymbol{y}^{(M)}\sim\mathcal{C}\mathcal{N}(&\sqrt{P N^2 C^4 } x \; \text{E}[ \boldsymbol{G}^{(M)} \boldsymbol{v}^T] + \text{E}[\boldsymbol{n}^{(M)}],  \\ &{P N^2 C^4 } ||x||^2\; \text{Cov}[ \boldsymbol{G}^{(M)} \boldsymbol{v}^T]+\text{Cov}[\boldsymbol{n}^{(M)}]).
\end{align}
\noindent
Recall that $\boldsymbol{n}^{(M)}\sim \mathcal{C}\mathcal{N}(0,N_0 \boldsymbol{I}_M)$, where $\boldsymbol{I}_M$ is the $M\times M$ identity matrix. Also recall that the pilot signal $x$ has unit energy, i.e., $||x||_2=1$. For the signal component, as each of the path coefficients $\alpha_l$ have zero mean we can write $\text{E}[ \boldsymbol{G}^{(M)} \boldsymbol{v}^T]  = 0$ and
\begin{align}
\label{E_h_v_2}
    \text{Cov}[ \boldsymbol{G}^{(M)} \boldsymbol{v}^T]&=\text{E}[ \boldsymbol{G}^{(M)} \boldsymbol{v}^T ( \boldsymbol{G}^{(M)} \boldsymbol{v}^T)^H] \\ &= N^2 C^4
      \boldsymbol{G}^{(M)} \text{E}[ \boldsymbol{v}^T \boldsymbol{v}] (\boldsymbol{G}^{(M)})^H.
\end{align}
\noindent
By defining a binary version of $\boldsymbol{v}$, denoted by $\boldsymbol{\bar{v}}$, with elements defined by
    \begin{equation}
    \label{v_bin}
     \bar{v}_{d} = \begin{cases}
    1, \text{ if } ||v_{d}||_2>0,\; \forall \; d=1,...,K^2  \; ; \\
    0, \text{ otherwise} \\
    \end{cases}
    \end{equation}
\noindent
we can separate the AOD/AOA information in $\bar{\boldsymbol{{v}}}$ from each of the path coefficient variances, $P_R$. As each path coefficient has variance $E[\alpha_l \alpha^*_l]=P_R, \; \forall \; l$, this then gives $\text{E}[\boldsymbol{v}^T \boldsymbol{v}] = P_R \boldsymbol{\bar{v}}^T \boldsymbol{\bar{v}} $. We can then re-write the distribution of $\boldsymbol{y}^{(M)}$ as
\begin{align}
\label{y_dist_K_v}
    \boldsymbol{y}^{(M)}\sim\mathcal{C}\mathcal{N}(0,\boldsymbol{\Sigma_v})
\end{align}
where
\begin{align}
\label{K_v}
    \boldsymbol{\Sigma_v} = P N^2 C^4 P_R \;
      \boldsymbol{G}^{(M)} \boldsymbol{\bar{v}}^T \boldsymbol{\bar{v}} (\boldsymbol{G}^{(M)})^{H} + N_0 \boldsymbol{I}_M.
\end{align}
\noindent
It can now be seen that $\boldsymbol{y}^{(M)}$ follows a zero mean,  circularly symmetric complex Gaussian (CSCG) distribution with corresponding probability density function (PDF) defined as \cite{gallager2008circularly}
\begin{align}
\label{f_pdf}
    f(\boldsymbol{y}^{(M)}|\boldsymbol{\bar{v}},\boldsymbol{G}^{(M)}) = \frac{1}{\pi^M\text{det}(\boldsymbol{\Sigma_v})} \text{exp}(-(\boldsymbol{y}^{(M)})^H \boldsymbol{\Sigma}^{-1}_v {\boldsymbol{y}^{(M)}}).
\end{align}
\noindent
Now let us find the conditional probability of $\boldsymbol{\bar{v}}$, given the receive measurement vector $\boldsymbol{y}^{(M)}$ and knowledge of $\boldsymbol{G}^{(M)}$, denoted by $p(\boldsymbol{\bar{v}}|\boldsymbol{y}^{(M)},\boldsymbol{G}^{(M)})$. Define $\mathcal{V}$ as the set of all possible binary  channel realizations such that $\boldsymbol{\bar{v}} \in \mathcal{V}$. We also define $|\mathcal{V}|$ to represent the cardinality of this set. Following the principle of maximum likelihood detection and based on Bayes rule \cite{bock1981marginal}, we can express the probability of $\boldsymbol{\bar{v}}$ for all possible $\boldsymbol{\bar{v}}\in \mathcal{V}$ as
\begin{align}
\label{prob}
   p(\boldsymbol{\bar{v}}|\boldsymbol{y}^{(M)},\boldsymbol{G}^{(M)}) = \frac{f(\boldsymbol{y}^{(M)}|\boldsymbol{{\bar{v}}},\boldsymbol{G}^{(M)})p(\boldsymbol{{\bar{v}}})}{p(\boldsymbol{y}^{(M)}|\boldsymbol{G}^{(M)})}
\end{align}
\noindent
where the term
\begin{align}
\label{prob_sum}
p(\boldsymbol{y}^{(M)}|\boldsymbol{G}^{(M)}) = \sum\limits_{\boldsymbol{\bar{v}}\in \mathcal{V}}  f(\boldsymbol{y}^{(M)}|\boldsymbol{{\bar{v}}},\boldsymbol{G}^{(M)}) p(\boldsymbol{{\bar{v}}})
\end{align}
\noindent
is independent of a particular channel realization. We assume that each channel realization is equiprobable, therefore
\begin{align}
\label{p_v}
p( \boldsymbol{{\bar{v}}} ) =  \frac{1}{|\mathcal{V}|},\; \forall \; \boldsymbol{\bar{v}} \in \mathcal{V}.
\end{align}
\noindent
We then denote the probability that the $d$th element of $\boldsymbol{\bar{v}}$ has a path by $p({\bar{v}}_{d}=1), \; \forall \; d=1,...,K^2$. We can express this probability as the sum of all $p(\boldsymbol{\bar{v}}|\boldsymbol{y}^{(M)},\boldsymbol{G}^{(M)})$ in (\ref{prob}) in which ${\bar{v}}_{d}=1$ by
\begin{align}
\label{v_soft}
p({\bar{v}}_{d}=1|\boldsymbol{y}^{(M)},\boldsymbol{G}^{(M)}) = \sum \limits_{\begin{array}{c} \bar{ \boldsymbol{v}} \in \mathcal{V} \\
\bar{\boldsymbol{v}}_d =1  \end{array}} p(\boldsymbol{\bar{v}}|\boldsymbol{y}^{(M)},\boldsymbol{G}^{(M)}).
\end{align}
\noindent
Following the maximum likelihood approach we then find the most likely sub-range combination by
\begin{align}
\label{d_est}
{\hat{d}}=\underset{{d}}{\operatorname{argmax }} [p({\bar{v}}_{d}=1|\boldsymbol{y}^{(M)},\boldsymbol{G}^{(M)})].
\end{align}
\noindent
Finally by finding the most likely transmitter and receiver subranges through
\begin{align}
\label{k_est}
\hat{k}_t= \Big{\lceil}\frac{\hat{d}}{K}\Big{\rceil},\; \hat{k}_r=\hat{d}-K(\hat{k}_t-1)
\end{align}
\noindent
we can reduce the ranges of possible AOD and AOA to, respectively, the $\hat{k}_t$th transmit and $\hat{k}_r$th receive angular sub-ranges. Each of these two sub-ranges will be further divided into another $K$ sub-ranges for the channel estimation in the next stage.

\subsection{Multi-stage Generalization}

In general the proposed channel estimation algorithm works in a similar multi-stage manner as that in \cite{rheath}, requiring $S=\lceil \text{log}_KN \rceil$ stages. {\color{\col}We show a high level overview of this process in  Fig. \ref{high_level}}. In the $s$th stage, we initially divide the possible AOA angular space into $K$ non-overlapped sub-ranges $\mathcal{S}^{(s)}_{r,1}, \mathcal{S}^{(s)}_{r,2}, \dots, \mathcal{S}^{(s)}_{r,K}$ and divide the possible AOD angular space into $\mathcal{S}^{(s)}_{t,1}, \mathcal{S}^{(s)}_{t,2}, \cdots, \mathcal{S}^{(s)}_{t,K}$. Then only $M$ overlapped beam pattern pairs will be designed at the transmitter and receiver to cover these $K$ sub-ranges. The designed $M$ beam patterns are characterized by the $M \times K$ beam pattern design matrices $\boldsymbol{B}_{T}^{(s,M)}$ and $\boldsymbol{B}_{R}^{(s,M)}$. These should be generated to maximize the minimum Euclidean distance between the columns of the corresponding generator matrix $\boldsymbol{G}^{(s,M)}$.

Given $\boldsymbol{B}_{T}^{(s,M)}$ and $\boldsymbol{B}_{R}^{(s,M)}$, we can then generate both the transmit beamforming vectors $\{\boldsymbol{f}^{(s)}_{m}\}$ and receive beamforming vectors $\{\boldsymbol{w}^{(s)}_{m}\}$ respectively in the same way as in (\ref{f_design})-(\ref{w_design}). For example, to generate ${\boldsymbol{f}^{(s)}_{m}}$, the corresponding vector $\boldsymbol{z}_{T}^{(m)}$ in (\ref{matrix_eq_1}), which is redefined as $\boldsymbol{z}_{T}^{(s,m)}$ for rigorousness, should be designed such that its $i$th entry, denoted by $[\boldsymbol{z}_{T}^{(s,m)}]_i, \; \forall \; i=1,...,N$, satisfies
    \begin{align}
    \label{desired_output_gen}
    &[\boldsymbol{z}_{T}^{(s,m)}]_i\!= \begin{cases}
    C_s {b}^{(m,k_t)}_{T} ,\!&\text{\!if $\frac{\pi i}{N} \in {\mathcal{S}_{t,k_t}^{(s)}, }\;\exists \;k_t \in \{1,\cdots,K \} ;$\!}  \\
    $0$,\!&\text{\!if $\frac{\pi i}{N} \notin {\mathcal{S}_{t,k_t}^{(s)}, } \;\forall \; k_t \in \{1,\cdots,K \} $\!} \!
    \end{cases}
    \end{align}
%
\begin{figure}[!t]
\removelatexerror
\begin{algorithm}[H]
\label{alg1}
\caption{Fast channel estimation (FCE) algorithm for mmWave channels using overlapped beam patterns}
{\fontsize{10}{10}\selectfont
$\mathbf{Input:}$ Transmitter and receiver both know, $N$, $K$\\ and have $\boldsymbol{B}_T^{(s,M)}$, $\boldsymbol{B}_R^{(s,M)}$. \vspace{2pt}  \\
$\mathbf{Initialization:}$ Set initial sub-ranges,\
$\mathcal{S}_{t,k}^{(1)},\mathcal{S}_{r,k}^{(1)} ,\;\forall\ \;k=1,\cdots,K$ \vspace{2pt}  \\	
   \For( \emph{}){$s\leq S $}
   {
   	   \vspace{2pt}  // Calculate: \vspace{2pt}  \\
   	   $\{\boldsymbol{f}^{(s)}_m\}$ based on $\mathcal{S}_{t,k}^{(s)},\;\forall \;k=1,\cdots,K$ and $\boldsymbol{B}_T^{(s,M)}$ \vspace{2pt}  \\
   	   $\{\boldsymbol{w}^{(s)}_m\}$ based on $\mathcal{S}_{r,k}^{(s)},\;\forall \;k=1,\cdots,K$ and $\boldsymbol{B}_R^{(s,M)}$ \vspace{2pt}  \\	
       \For( \emph{}){$m=1$ to $M$}
        {
          Transmitter transmits using $\boldsymbol{f}^{(s)}_{m}$ \vspace{2pt}  \\
          Receiver measures using $\boldsymbol{w}^{(s)}_{m}$ \vspace{2pt}  \\
        }
		// After $M$ measurements: \vspace{4pt}	\\	
		$\boldsymbol{y}^{(s,M)} = \sqrt{P} x \boldsymbol{h}_v^{(s,M)} +\boldsymbol{n}^{(s,M)}$  \vspace{2pt}  \\
		${\hat{d}^{(s)}} = \underset{{d}}{\operatorname{argmax }} [p({\bar{v}}_{d}=1|\boldsymbol{y}^{(s,M)},\boldsymbol{G}^{(s,M)})]$	\vspace{2pt}  \\
		$\hat{k}_t^{(s)}=\Big{\lceil}\frac{\hat{d}^{(s)}}{K}\Big{\rceil},\; \hat{k}_r^{(s)}=\hat{d}^{(s)}-K(\hat{k}^{(s)}_t-1)$ \\
		// Refine sub-ranges based on $\hat{k}_t^{(s)}$ and $\hat{k}_r^{(s)}$ \\
		$\mathcal{S}_{t,k}^{(s+1)},\mathcal{S}_{r,k}^{(s+1)} ,\;\forall\ \;k=1,\cdots,K$.
   }
$\mathbf{Output:}$
   $\hat{\phi^t}=\frac{\pi}{N} \sum\limits_{s=1}^S (\hat{k}^{(s)}_t-1)K^{S-s}  ,\;\hat{\phi^r}= \frac{\pi}{N} \sum\limits_{s=1}^S (\hat{k}^{(s)}_r-1)K^{S-s}$
      $\hat{\alpha}= P_R \boldsymbol{\hat{r}}^H ( \boldsymbol{\hat{r}} P_R \boldsymbol{\hat{r}}^H + N_0 \boldsymbol{I}_{|\boldsymbol{\hat{r}}|})^{-1} \boldsymbol{\boldsymbol{r}}$
}
\end{algorithm}
\end{figure}

\noindent
where $C_s$ is a scalar constant for the $s$th stage to guarantee that $\boldsymbol{f}^{{(s)}}_{{m}}$ satisfies $||\boldsymbol{f}_m^{(s)}||_2=1$. Physically, $[\boldsymbol{z}_{T}^{(s,m)}]_i$ describes the desired beam pattern amplitude at angle $\frac{\pi i}{N}$ when $\boldsymbol{f}^{{(s)}}_{{m}}$ is used. Each receive beamforming vector ${\boldsymbol{w}^{(s)}_{m}}$ can be designed in the same way.

The channel output on the $s$th estimation stage can then be obtained after $M$ time slots by
\begin{align}
\label{y_s}
\boldsymbol{y}^{(s,M)} =& \sqrt{P_s} x \boldsymbol{h}_v^{(s,M)} + \boldsymbol{n}^{(s,M)}
\end{align}
where
\begin{align}
\label{h_v_s_m}
\boldsymbol{h}_v^{(s,M)}  =   \left[\begin{array}{ccc}
   	( \boldsymbol{w}_1^{(s)})^H \boldsymbol{H} \boldsymbol{f}_1^{(s)}  \\
   	( \boldsymbol{w}_2^{(s)})^H \boldsymbol{H} \boldsymbol{f}_2^{(s)}  \\
   	\vdots												\\ 	
   	( \boldsymbol{w}_M^{(s)})^H \boldsymbol{H} \boldsymbol{f}_M^{(s)}
    \end{array}\right]
\end{align}
\noindent
and $P_s$ denotes the transmit power of the pilot signal in the $s$th stage. Similar to that in \cite{rheath}, we prefer that all the stages have an equal probability of failure, indicating that we should allocate power among stages inversely proportional to the beamforming gains of these beam patterns, i.e.,
    \begin{equation}
    \label{p_s}
    {P_{s}}=\frac{{P_T}}{C_{s}^4}, \; \forall \; s=1,2,\cdots,S
    \end{equation}
\noindent
where $P_T$ is a constant. Similar to (\ref{d_est}) we then find the most likely sub-range combination of the $s$th stage by
\begin{align}
\label{d_est_stage}
{\hat{d}^{(s)}} = \underset{{d}}{\operatorname{argmax }} [p({\bar{v}}_{d}=1|\boldsymbol{y}^{(s,M)},\boldsymbol{G}^{(s,M)})]
\end{align}
\noindent
with the corresponding most likely transmitter and receiver sub-ranges given by
\begin{align}
\label{k_est_s}
\hat{k}_t^{(s)}= \Big{\lceil}\frac{\hat{d}^{(s)}}{K}\Big{\rceil},\; \hat{k}_r^{(s)}=\hat{d}^{(s)}-K(\hat{k}_t^{(s)}-1).
\end{align}
\noindent
The selected sub-ranges, $\mathcal{S}_{t,\hat{k}_t^{(s)}}^{(s)}$ and $\mathcal{S}_{r,\hat{k}_r^{(s)}}^{(s)}$ are then used for the channel estimation in the next stage.
This process continues until the minimum angle resolution $\frac{\pi}{N}$ is reached requiring $S=\lceil \text{log}_K(N) \rceil$ stages. It is worth pointing out that although the proposed algorithms are elaborated based on the estimation process of a single path, their implementation in multi-path scenario is actually feasible by following the same procedure as in [\citenum{rheath}, Algorithm 2]. More specifically, multiple paths are estimated sequentially, with the first path being estimated using the multi-stage algorithms described above. Subsequent paths can then be found by returning to the first stage and repeating the estimation. Moreover, in each stage's measurements, the expected contributions from all previously estimated paths can be subtracted to reveal new paths.

\subsection{Estimation of the Fading Coefficient} \label{sssec:MMSE}
Once all estimation stages described in the previous subsection have been performed, we estimate the identified path fading coefficient $\alpha$. In \cite{rheath}, the value of ${\alpha}$ was estimated based on the measurement of the final stage only. To improve the estimation accuracy, we estimate ${\alpha}$ by using all measurements in all stages of the algorithm. Denote by $\boldsymbol{r}$ and $\boldsymbol{\hat{r}}$, respectively, the vector of all received measurements and the vector of their estimates such that
    \begin{equation}
    \label{r_corr}
	\boldsymbol{r} =  \left[\begin{array}{ccc}
   	\boldsymbol{y}^{(1,M)}  \\
   	\vdots					\\ 	
   	\boldsymbol{y}^{(S,M)}
    \end{array}\right],\;
	\boldsymbol{\hat{r}} =  \left[\begin{array}{ccc}
   	x\sqrt{P_1}N C_1^2 \boldsymbol{G}^{(1,M)}_{\hat{d}^{(1)}}  \\
   	\vdots					\\ 	
    x\sqrt{P_S}N C_S^2 \boldsymbol{G}^{(S,M)}_{\hat{d}^{(S)}}
    \end{array}\right]
    \end{equation}
\noindent
where $\boldsymbol{G}^{(s,M)}_i$ denotes the $i$th column of $\boldsymbol{G}^{(s,M)}$. Provided that the AOD/AOA estimation is correct in each stage, we can write
    \begin{equation}
    \label{r_bold}
	\boldsymbol{{r}} = \boldsymbol{\hat{r}} \alpha + \boldsymbol{w}
    \end{equation}
\noindent
where $\boldsymbol{w}$ is the $SM\times 1$ vector of corresponding noise terms. In the case where the AOD/AOA estimation is incorrect, the estimation of the fading coefficient is not important as there will be a beam misalignment between the transmitter and receiver. Following the LMMSE principle \cite{scharf1991statistical}, we can then estimate the fading coefficient $\alpha$ as
\begin{equation}
\label{alpha_est}
   \hat{\alpha}= P_R \boldsymbol{\hat{r}}^H ( \boldsymbol{\hat{r}} P_R \boldsymbol{\hat{r}}^H + N_0 \boldsymbol{I}_{|\boldsymbol{\hat{r}}|})^{-1} \boldsymbol{\boldsymbol{r}}
\end{equation}
\noindent
where $\boldsymbol{I}_{|\boldsymbol{\hat{r}}|}$ is an ${|\boldsymbol{\hat{r}}|} \times {|\boldsymbol{\hat{r}}|}$ identity matrix. Now we can formally describe the proposed fast channel estimation algorithm using overlapped beam patterns in {\bf{Algorithm \ref{alg1}}}.

\noindent { \textit{\textbf{Remark 1.}}} It can be seen that, compared with the channel estimation algorithm in \cite{rheath} with the same value of $K$, our proposed algorithm also requires $S=\lceil \text{log}_K N \rceil$ stages, but the number of measurement time slots required in each stage reduces to $M$, instead of $K^2$. In general, this yields a $\frac{K^2}{M}$ reduction in measurement time slots. For the example of $K=3$ discussed earlier with $M=4$, a $\frac{K^2}{M^2}=225\%$ increase in estimation rate can be achieved.

\section{Rate Adaptive Channel Estimation Algorithm}
The proposed channel estimation scheme explained in the previous section uses a fixed $\boldsymbol{G}^{(s,M)}$ where the detector is forced to make a decision after $M$ measurements, irrespective of what the computed probability $p({\bar{v}}_{\hat{d}^{(s)}}=1|\boldsymbol{y}^{(s,M)},\boldsymbol{G}^{(s,M)})$ may be. Leveraging the detection method developed in the previous section, we now propose a novel rate-adaptive channel estimation (RACE) algorithm.

We first introduce a target maximum probability of estimation error (PEE), denoted by $\Gamma$. The basic principle of the RACE algorithm is that after the $M$ initial measurements are completed in any given stage, if the most likely sub-range combination probability does not satisfy $p({\bar{v}}_{\hat{d}^{(s)}}=1|\boldsymbol{y}^{(s,M)},\boldsymbol{G}^{(s,M)})>(1-\Gamma)$, then additional measurements will be performed. To this end, the receiver will feedback the current most likely transmit sub-range, $\hat{k}_t^{(s)}$, and also the information indicating whether more measurements are required or not.

\begin{figure}[!t]
\removelatexerror
\begin{algorithm}[H]
\label{alg2}
\caption{Rate-adaptive channel estimation (RACE) algorithm for mmWave channels}
{\fontsize{10}{10}\selectfont
$\mathbf{Input:}$ Transmitter and receiver both know, $N$, $K$\\ and have $\boldsymbol{B}_T^{(s,M)}$, $\boldsymbol{B}_R^{(s,M)}$. \vspace{2pt}  \\
$\mathbf{Initialization:}$ Set initial sub-ranges,\
$\mathcal{S}_{t,k}^{(1)},\mathcal{S}_{r,k}^{(1)}, \;\forall\ \;k=1,\cdots,K$ \vspace{2pt}  \\	
   \For( \emph{}){$s\leq S $}
   {
   	   \vspace{4pt} // Calculate: \vspace{2pt}  \\
   	   $\{\boldsymbol{f}^{(s)}_m\}$ based on $\mathcal{S}_{t,k}^{(s)}, \;\forall \;k=1,\cdots,K$ and $\boldsymbol{B}_T^{(s,M)}$ \vspace{2pt}  \\
   	   $\{\boldsymbol{w}^{(s)}_m\}$ based on $\mathcal{S}_{r,k}^{(s)},\;\forall \;k=1,\cdots,K$ and $\boldsymbol{B}_R^{(s,M)}$ \vspace{2pt}  \\	
       \For( \emph{}){$m=1$ to $M$}
        {
          Transmitter transmits using $\boldsymbol{f}^{(s)}_{m}$ \vspace{2pt}  \\
          Receiver measures using $\boldsymbol{w}^{(s)}_{m}$ \vspace{2pt}  \\
        }
		\vspace{4pt} // After $M$ measurements: \vspace{4pt} \\
		$\boldsymbol{y}^{(s,M)} = \sqrt{P} x \boldsymbol{h}_v^{(s,M)} +\boldsymbol{n}^{(s,M)}$  \vspace{2pt}  \\
				${\hat{d}^{(s)}} = \underset{{d}}{\operatorname{argmax }} [p({\bar{v}}_{d}=1|\boldsymbol{y}^{(s,M)},\boldsymbol{G}^{(s,M)})]$	\vspace{2pt}  \\
		$\hat{k}_t^{(s)}=\Big{\lceil}\frac{\hat{d}^{(s)}}{K}\Big{\rceil},\; \hat{k}_r^{(s)}=\hat{d}^{(s)}-K(\hat{k}^{(s)}_t-1)$ \vspace{2pt} \\
		 \vspace{2pt} // Carry out additional measurements if required: \vspace{2pt}  \\
		$R=0$\\		
		\While( \emph{}){$ p({\bar{v}}_{\hat{d}^{(s)}}=1|\boldsymbol{y}^{(s,M+R)},\boldsymbol{G}^{(s,M+R)})< (1-\Gamma)  $\\$ {\text{ \bf{and} }}(M+R<M_{max}) $}{
			\vspace{4pt}
			$R=R+1$ // Increment re-measurement index \\
			Transmitter transmits with: \vspace{4pt}	\\
			$\boldsymbol{f}_{M+R}^{(s)}$ based on $\mathcal{S}_{t,k}^{(s)},\;\forall \;k$ and $\overrightarrow{b}_{\hat{k}_t^{(s)}}^K\vspace{2pt}$  \\
			Receiver measures with: \vspace{4pt}	\\
			$\boldsymbol{w}_{M+R}^{(s)}$ based on $\mathcal{S}_{r,k}^{(s)},\;\forall \;k$ and $\overrightarrow{b}_{\hat{k}_r^{(s)}}^K\vspace{2pt}$  \\
			Update: \vspace{4pt}	\\
		${\hat{d}^{(s)}} = \underset{{d}}{\operatorname{argmax }} [p({\bar{v}}_{d}=1|\boldsymbol{y}^{(s,M+R)},\boldsymbol{G}^{(s,M+R)})]$	\vspace{2pt}  \\
		$\hat{k}_t^{(s)}=\Big{\lceil}\frac{\hat{d}^{(s)}}{K}\Big{\rceil},\; \hat{k}_r^{(s)}=\hat{d}^{(s)}-K(\hat{k}^{(s)}_t-1)$			
		}
		\vspace{4pt} // Refine sub-ranges based on $\hat{k}_t^{(s)}$ and $\hat{k}_r^{(s)}$ \\
		$\mathcal{S}_{t,k}^{(s+1)},\mathcal{S}_{r,k}^{(s+1)} ,\;\forall\ \;k=1,\cdots,K$.
   }
$\mathbf{Output:}$
   $\hat{\phi^t}=\frac{\pi}{N} \sum\limits_{s=1}^S (\hat{k}^{(s)}_t-1)K^{S-s}  ,\;\hat{\phi^r}= \frac{\pi}{N} \sum\limits_{s=1}^S (\hat{k}^{(s)}_r-1)K^{S-s}$
      $\hat{\alpha}= P_R \boldsymbol{\hat{r}}^H ( \boldsymbol{\hat{r}} P_R \boldsymbol{\hat{r}}^H + N_0 \boldsymbol{I}_{|\boldsymbol{\hat{r}}|})^{-1} \boldsymbol{\boldsymbol{r}}$
}
\end{algorithm}
\end{figure}

Note that the non-overlapped algorithm proposed in \cite{rheath} also feeds back a sub-range number to the transmitter, however we include an additional bit corresponding to whether more measurements in this stage are required or not. This feedback only requires $\lceil\text{log}_2(K)+1\rceil$ bits, which will be shown to be negligible at high signal-to-noise ratio (SNR) as the average number of additional measurements required is close to zero.

If the specified probability threshold was not met after the $M$th measurement, instead of further dividing sub-ranges corresponding to $\hat{k}_t^{(s)}$ and $\hat{k}_r^{(s)}$ at the transmitter and receiver, additional measurement will be further applied to the current stage. {\color{\col}Since the most likely sub-range combination has been determined based on the previous $M$ overlapped measurements, there is no motivation to measure on multiple sub-range combinations in the new measurement. In other words, the best strategy of the system for the subsequent measurement is to measure the most likely sub-range combination only. It is also important to note, however, that the beam pattern combination used in the new measurement will still be overlapped with previous measurements already taken on this most likely sub-range combination.} The beamforming vectors associated with the new measurement are designed to correspond to a newly added row to each of $\boldsymbol{B}_T^{(s,M)}$ and $\boldsymbol{B}_R^{(s,M)}$ yielding $\boldsymbol{B}_T^{(s,M+1)}$ and $\boldsymbol{B}_R^{(s,M+1)}$, respectively. We define $\overrightarrow{b}_i^K$ as a $1 \times K$ sparse binary row vector with a single $1$ in its $i$th entry and $0$ otherwise such that $\overrightarrow{b}_i^K = [\boldsymbol{0}_{i-1}, 1, \boldsymbol{0}_{K-i}]$ where $\boldsymbol{0}_i$ is a $1\times i$ all zeros row vector. The updated beam pattern design matrices $\boldsymbol{B}_T^{(s,M+1)}$ and $\boldsymbol{B}_R^{(s,M+1)}$ can then be expressed as

    \begin{equation}
    \label{B_general_staged}
    \boldsymbol{B}_T^{(s,M+1)}=\left[\begin{array}{ccc}
	\boldsymbol{B}_T^{(s,M)} \\
	\overrightarrow{b}_{\hat{k}_t^{(s)}}^K
    \end{array}\right],\;
    \boldsymbol{B}_R^{(s,M+1)}=\left[\begin{array}{ccc}
	\boldsymbol{B}_R^{(s,M)} \\
	\overrightarrow{b}_{\hat{k}_r^{(s)}}^K
    \end{array}\right].
    \end{equation}
\noindent
The $(M+1)$th transmit and receive beamforming vectors can then be calculated from the new row in these matrices and be used to measure the channel, obtaining $\boldsymbol{y}^{(s,M+1)}$. The updated generator matrix corresponding to (\ref{B_general_staged}) can be described by
    \begin{equation}
    \label{G_general_staged}
    \boldsymbol{G}^{(s,M+1)}= \boldsymbol{B}_T^{(s,M+1)} \odot  \boldsymbol{B}_T^{(s,M+1)} = \left[\begin{array}{ccc}
    \boldsymbol{G}^{(s,M)} \\
    \overrightarrow{b}^{K^2}_{\hat{d}^{(s)}}
    \end{array}\right].
    \end{equation}
The estimation parameters can then be updated based on the updated $\boldsymbol{G}^{(s,M+1)}$ and $\boldsymbol{y}^{(s,M+1)}$ and the ML detector developed in the previous section. We can generalize this to $\boldsymbol{G}^{(s,M+R)}$ where $R=0,1,...$ indexes the additional measurements.

The proposed RACE algorithm is formally described in {\bf{Algorithm \ref{alg2}}}. As can be seen from the algorithm description, this process then repeats until either the threshold condition has been met (i.e., $p({\bar{v}}_{\hat{d}}=1|\boldsymbol{y}^{(s,M)},\boldsymbol{G}^{(s,M)})>(1-\Gamma)$) or a maximum number of measurements, denoted by $M_{max}$, has been reached (i.e., $M+R=M_{max}$). Under fading channel conditions, there always exists a non-zero probability of an ‘outage’ occurring when the path coefficient is close to zero. By imposing the upper limit to the number of measurements we reduce the time and energy expended in this case. As the RACE algorithm is able to compute the probability of successful estimation during the estimation process, it can minimize the number of measurements required for successful estimation and therefore reduce the associated energy.
\section{Performance Analysis}
\label{sec:Probability}
In this section, we focus on the performance analysis of the proposed algorithms when a single path is present between the transmitter and receiver. Specifically, we derive two analytical expressions. The first is an expression for the PEE (i.e., an incorrect estimation of the AOA or AOD), whereas the second expression is for the minimum energy-to-noise ratio required by the RACE algorithm for a specified average number of measurement, $M+R$.

    \begin{figure*}[!t]
    \centering
    \subfigure[]{\includegraphics[width=3.5in,trim={1.5cm 7cm 1cm 8.5cm},clip]{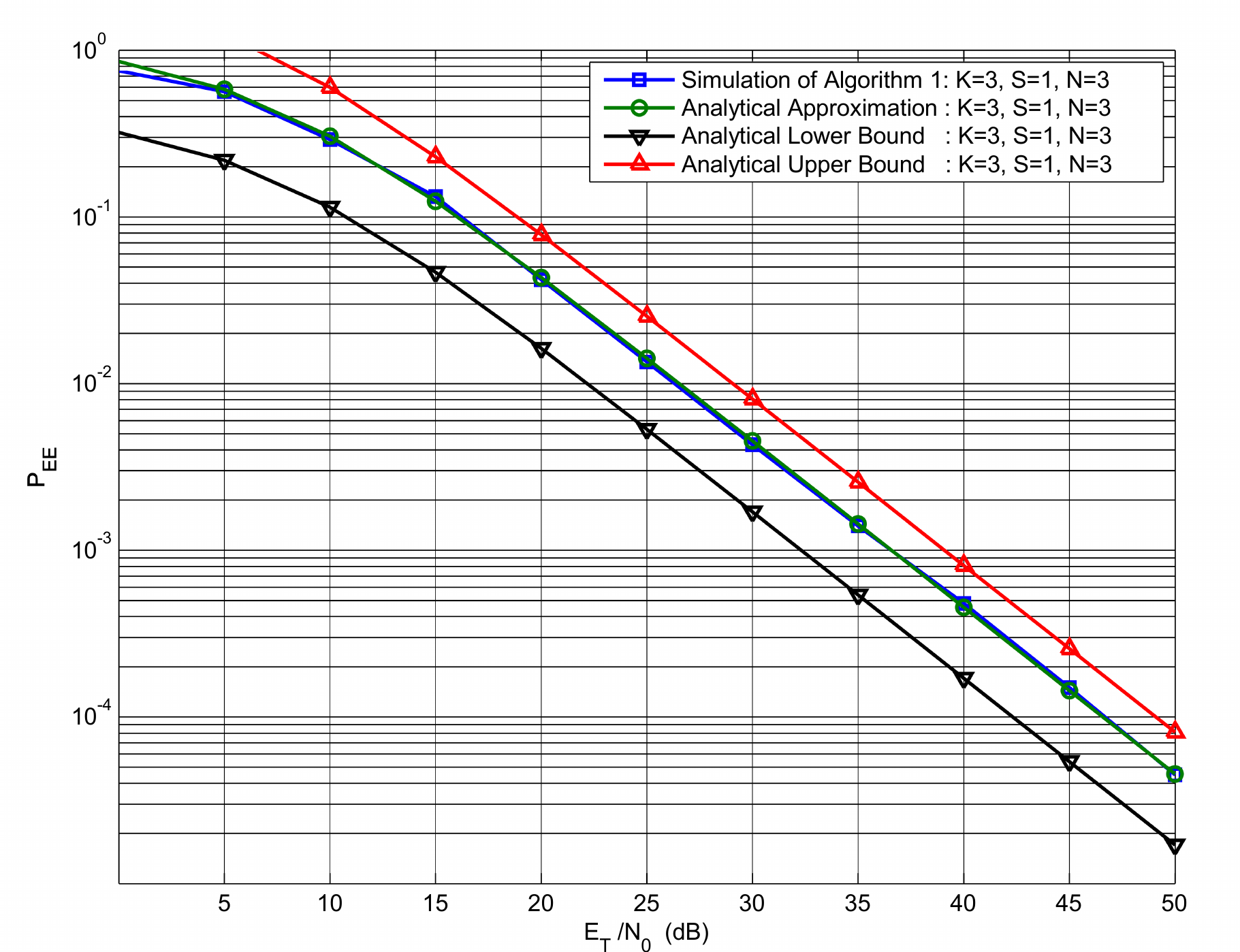}}
    \subfigure[]{\includegraphics[width=3.5in,trim={1.5cm 7cm 1cm 8.5cm},clip]{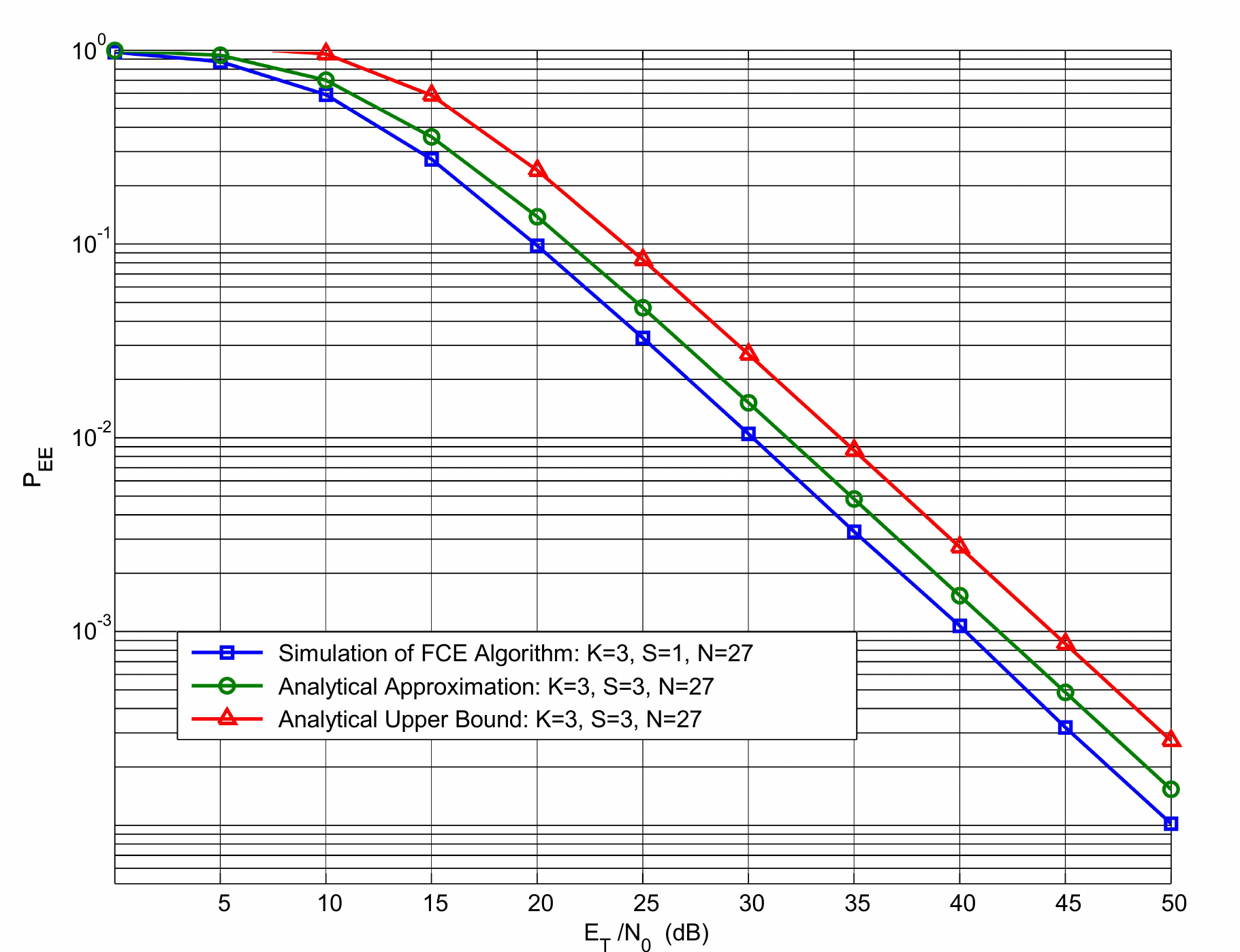}}
    \caption{Comparison of the AOA/AOD probability of estimation error numerical results with the derived analytical expression.  Fig. (a) uses shows the PEE for a single stage system, whereas Fig. (b) shows a system with $S=3$. Both systems use $P_R=1$ and are compared to the total energy per noise ratio, calculated by $E_T=M\sum_{s=1}^SP_s$. The Analytical Approximation, Lower Bound and Upper Bound in (a) refer to, (\ref{P_approx}),  (\ref{P_LB}) and (\ref{P_EE_3}), respectively. In (b), the Analytical Approximation refers to the substitution of (\ref{P_approx}) into (\ref{P_ov_2}) where as the Analytical Upper Bound refers to the substitution of (\ref{P_EE_3}) into (\ref{P_ov_2}).}
    \label{analytical_EER}
    \end{figure*}

\subsection{Probability of AOD/AOA Estimation Error}

As the RACE algorithm seeks to achieve a fixed error rate by means of an adaptive $\boldsymbol{G}^{(s,M+R)}$, we first need to find the PEE for a fixed generator matrix, $\boldsymbol{G}^{(s,M+R)}, \; \forall \; s=1,...,S$. Note that this is also the case in Algorithm \ref{alg1} with a fixed generator matrix, $\boldsymbol{G}^{(s,M)}, \; \forall \; s=1,...,S$. We begin by defining the PEE in a given stage $s$, assuming all previous stages have been correct, as
\begin{align}
\label{P_EE}
p_{EE|\boldsymbol{G}^{(s,M+R)},\bar{\boldsymbol{v}}^{(s)}} = p(\bar{\boldsymbol{v}}^{(s)} \neq \hat{\boldsymbol{v}}^{(s)} )
\end{align}
%
where the term $p(\bar{\boldsymbol{v}}^{(s)} \neq \hat{\boldsymbol{v}}^{(s)} )$ represents the probability that the estimated AOD/AOA information $\hat{\boldsymbol{v}}^{(s)}$ is not equal to the physical channel, $\bar{\boldsymbol{v}}^{(s)}$.

If we treat the channel vector $\bar{\boldsymbol{v}}^{(s)}$ as information being encoded by the generator matrix $\boldsymbol{G}^{(s,M+R)}$, we can model this estimation error event as a maximum likelihood decoding error. We can then say that $\bar{\boldsymbol{v}}^{(s)} \neq \hat{\boldsymbol{v}}^{(s)}$ occurs when the received measurement vector $\boldsymbol{y}^{(s)}=\alpha x \sqrt{P_sN^2C_s^4} \boldsymbol{G}^{(s,M+R)} (\bar{\boldsymbol{v}}^{(s)})^T +\boldsymbol{n}^{(s)}$, is incorrectly 'decoded' to another received measurement vector $\boldsymbol{y}'^{(s)}=\alpha x \sqrt{P_sN^2C_s^4} \boldsymbol{G}^{(s,M+R)} \bar{\boldsymbol{v}}'^T,\; \forall \; \bar{\boldsymbol{v}}' \in \mathcal{V} \; \text{and} \; \bar{\boldsymbol{v}}' \neq \bar{\boldsymbol{v}}^{(s)} $. We then express the probability of this event as
\begin{align}
\label{P_v}
p(\bar{ \boldsymbol{v} }^{(s)} \neq \hat{ \boldsymbol{v}}^{(s)} ) = \bigcup  \limits_{{\begin{array}{c} \bar{ \boldsymbol{v}}' \in \mathcal{V} \\
\bar{\boldsymbol{v}}' \neq  \bar{\boldsymbol{v}} \end{array} }} p(\boldsymbol{y}^{(s)} \rightarrow \boldsymbol{y}'^{(s)}).
\end{align}
As the decoding approach used in this paper is based upon maximum likelihood, we can calculate such an error probability from the Euclidean distance between the received measurement sets. With reference to \cite{stuber2011principles}, we can then express the pairwise error probability of this decoding error over a fading channel with coefficient $\alpha \sim \mathcal{C}\mathcal{N}(0,P_R)$ as
\begin{align}
\label{P_v_pairwise}
p(\boldsymbol{y}^{(s)}  \rightarrow \boldsymbol{y}'^{(s)} )&= \frac{1}{2} \Bigg[ 1-\sqrt{ \frac{\bar{\gamma} ^2}{\bar{\gamma} ^2+2} } \Bigg]
\end{align}
where
\begin{align}
\label{gamma}
\bar{\gamma} = \sqrt{\frac{P_s C_s^4 N^2 P_R}{ 2 N_0} } || {\boldsymbol{G}^{(s,M+R)}}(\bar{\boldsymbol{v}}^{(s)}-\bar{\boldsymbol{v}}')^T||
\end{align}
By substituting these terms into (\ref{P_EE}) and averaging over each $\bar{\boldsymbol{v}}^{(s)}$ we get

\begin{align}
p_{EE|\boldsymbol{G}^{(s,M+R)}} &=\sum \limits_{\bar{\boldsymbol{v}}  \in \mathcal{V}}  p(\boldsymbol{\bar{\boldsymbol{v}}})  p_{EE|\boldsymbol{G}^{(s,M+R)},\bar{\boldsymbol{v}}^{(s)}} \\
&=\sum \limits_{\bar{\boldsymbol{v}} \in \mathcal{V}}  \bigcup  \limits_{  {\begin{array}{c} \bar{\boldsymbol{v}}' \in \mathcal{V} \\
\bar{\boldsymbol{v}}' \neq  \bar{\boldsymbol{v}} \end{array} } }  \frac{p(\boldsymbol{\bar{\boldsymbol{v}}})}{2} \Bigg[ 1-\sqrt{ \frac{\bar{\gamma} ^2}{\bar{\gamma} ^2+2} } \Bigg].
\label{P_EE_2}
\end{align}
Expressing the exact probability of the union term in (\ref{P_EE_2}) is quite difficult due to complicated boundaries between each column in $\boldsymbol{G}^{(s,M+R)}$. This is a direct result of the generator matrix not containing a full set of normalized $(M+R)\times 1$ vectors. For example, (\ref{G_K_3_2}) does not contain the scaled versions of the vectors $[0,1,1,0]^T$  , $[1,1,0,1]^T$, $[1,0,1,1]^T$, $[1,1,1,0]^T$, $[0,1,1,1]^T$ or $[0,0,0,0]^T$. This is an inherent property of the generator matrix being constructed from two smaller matrices $\boldsymbol{B}_T^{(s,M)}$ and $\boldsymbol{B}_R^{(s,M)}$ that do not contain the all-zero column vector. We can, however, find an upper bound of expression in (\ref{P_EE_2}) by replacing the union term with a summation to get
\begin{align}
\label{P_EE_3}
p_{EE|\boldsymbol{G}^{(s,M+R)}} < \sum \limits_{\bar{\boldsymbol{v}} \in \mathcal{V}}  \sum  \limits_{  {\begin{array}{c} \bar{\boldsymbol{v}}' \in \mathcal{V} \\
\bar{\boldsymbol{v}}' \neq  \bar{\boldsymbol{v}} \end{array} } }  \frac{p(\boldsymbol{\bar{\boldsymbol{v}} })}{2} \Bigg[ 1-\sqrt{ \frac{\bar{\gamma} ^2}{\bar{\gamma} ^2+2} } \Bigg].
\end{align}
The resulting upper bound in (\ref{P_EE_3}) is found to be quite loose in most cases. Motivated by this, we find a much tighter approximation by only considering $\bar{\boldsymbol{v}}'$ that minimizes $|| {\boldsymbol{G}^{(s,M+R)}}(\bar{\boldsymbol{v}}^{(s)}-\bar{\boldsymbol{v}}')^T||$ (i.e., only considering spatially adjacent columns of $\boldsymbol{G}^{(s,M+R)}$). We define this minimum distance for each $\bar{\boldsymbol{v}}^{(s)}$ by
\begin{align}
\label{d_min}
d_{min}(\bar{\boldsymbol{v}}^{(s)})=\underset{   {\begin{array}{c} \bar{\boldsymbol{v}}' \in \mathcal{V} \\
\bar{\boldsymbol{v}}' \neq  \bar{\boldsymbol{v}} \end{array} } } {  \operatorname{min}} || {\boldsymbol{G}^{(s,M+R)}}(\bar{\boldsymbol{v}}^{(s)}-\bar{\boldsymbol{v}}')^T||
\end{align}
and the set of vectors causing it as
\begin{align}
\label{V_min_set}
\mathcal{V}_{min}(\bar{\boldsymbol{v}}^{(s)})=\Big{\{}\underset{   {\begin{array}{c} \bar{\boldsymbol{v}}' \in \mathcal{V}_{min} \\
\bar{\boldsymbol{v}}' \neq  \bar{\boldsymbol{v}} \end{array} }  } {  \operatorname{argmin}} || {\boldsymbol{G}^{(s,M+R)}}(\bar{\boldsymbol{v}}^{(s)}-\bar{\boldsymbol{v}}')^T||\Big{\}}.
\end{align}
We therefore approximate $p_{EE|\boldsymbol{G}^{(s,M+R)}}$ by
\begin{align}
\label{P_approx}
p_{EE|\boldsymbol{G}^{(s,M+R)}} \approx \;  \sum \limits_{\bar{\boldsymbol{v}} \in \mathcal{V}}  \sum  \limits_{  {\begin{array}{c} \bar{\boldsymbol{v}}' \in \mathcal{V}_{min} \\
\bar{\boldsymbol{v}}' \neq  \bar{\boldsymbol{v}} \end{array} } }  \frac{p(\boldsymbol{\bar{\boldsymbol{v}} })}{2} \Bigg[ 1-\sqrt{ \frac{\bar{\gamma} ^2}{\bar{\gamma} ^2+2} } \Bigg].
\end{align}
We also provide a greatly simplified lower bound for Algorithm \ref{alg1} when $R=0$, by only considering a single one of these $\bar{\boldsymbol{v}}'$ that cause a minimum distance. Recall that the generator matrix is designed such that all paths have the same minimum Euclidean distance, which we define here as $\sqrt{E_{min}}=d_{min}(\bar{\boldsymbol{v}}^{(s)}),\; \forall \; \bar{\boldsymbol{v}}^{(s)} $. By assuming each channel realization is equiprobable, this leads to
\begin{align}
\label{P_LB}
p_{EE|\boldsymbol{G}^{(s,M)}} > \frac{1}{2} \Bigg[ 1-\sqrt{ \frac{{{P_s C_s^4 N^2 P_R {E_{min}}} }}{{{P_s C_s^4 N^2 P_R {E_{min}}} }+4 N_0} } \; \Bigg]
\end{align}
Finally, we upper bound the probability of an error occurring in any of the $S$ stages as
\begin{align}
\label{P_ov}
p_{EE} &= 1 - \prod \limits_{s=1}^{S} (1-p_{EE|\boldsymbol{G}^{(s,M+R)}}) \\
              &< \; \sum \limits_{s=1}^{S} p_{EE|\boldsymbol{G}^{(s,M+R)}} \label{P_ov_2} \\
              &< \; \sum \limits_{s=1}^{S} \sum \limits_{\bar{\boldsymbol{v}} \in \mathcal{V}}  \sum  \limits_{  {\begin{array}{c} \bar{\boldsymbol{v}}' \in \mathcal{V} \\
\bar{\boldsymbol{v}}' \neq  \bar{\boldsymbol{v}}^{(s)} \end{array} } }  \frac{p(\boldsymbol{\bar{\boldsymbol{v}} })}{2} \Bigg[ 1-\sqrt{ \frac{\bar{\gamma} ^2}{\bar{\gamma} ^2+2} } \Bigg].  \label{P_ov_3}
\end{align}

In order to validate our analysis, Figs. \ref{analytical_EER} (a) and (b) compare the derived analytical expressions with their corresponding Monte Carlo simulations for a single stage system and three stage system, respectively. As we can observe from Fig. \ref{analytical_EER}(a), the given in (\ref{P_approx}) is actually quite tight. Focusing on the three stage system in Fig. \ref{analytical_EER}(b), we see that the substitution of the approximation in (\ref{P_approx}) into the upper bound (\ref{P_ov_2}) causes some disparity, however still provides a tighter approximation than (\ref{P_ov_3}). A lower bound is not presented in Fig. \ref{analytical_EER}(b) as the use of (\ref{P_LB}) in the upper bound (\ref{P_ov_2}) would not be mathematically correct.
\subsection{Minimum Energy and Time Requirement for the RACE Algorithm}
We now present an expression for the minimum signal energy-to-noise ratio required by the RACE algorithm for an average number of measurements in a given stage. From an information theory standpoint, we can describe the RACE algorithm as the information process that a $1\times K^2$ binary vector $\bar{\boldsymbol{v}}^{(s)}$ is encoded into $M+R$ symbols by the generator matrix $\boldsymbol{G}^{(s,M+R)}$. This information transfer has an equivalent modulation rate of $C=K^2/(M+R)$ information bits per measurement time slot duration. The Shannon-Hartley theorem \cite{shannon2001mathematical} tells us that the minimum received SNR required for error free detection at this rate is
\begin{align}
\label{shannon}
SNR_r&\geq 2^C-1 \nonumber \\
&= 2^{\frac{K^2}{M+R}}-1
\end{align}
where from (\ref{y_g}), for a given $\bar{\boldsymbol{v}}^{(s)}$, $\boldsymbol{G}^{(s,M+R)}$ and $\alpha$, we know the average received SNR can be expressed as
\begin{align}
\label{snr_r}
SNR_r = \frac{ ||\alpha||^2 P_s C_s^4 N^2 }{ N_0}  \frac{|| \boldsymbol{G}^{(s,M+R)}(\bar{\boldsymbol{v}}^{(s)})^T  ||^2}{(M+R)}.
\end{align}
Substituting (\ref{snr_r}) into (\ref{shannon}) gives
\begin{align}
\label{Rate_1}
 \frac{  ||\alpha||^2 P_s C_s^4 N^2}{ N_0}  \frac{|| \boldsymbol{G}^{(s,M+R)}(\bar{\boldsymbol{v}}^{(s)})^T  ||^2}{(M+R)} \geq 2^{\frac{K^2}{(M+R)}}-1.
\end{align}
By denoting the total energy used in stage $s$ by $E_s=P_s(M+R)$, we can re-write (\ref{Rate_1}) as
\begin{align}
\label{Rate_2}
 \frac{ E_s}{ N_0}   \geq \frac{(M+R)^2(2^{\frac{K^2}{(M+R)}}-1)}{ ||\alpha||^2 C_s^4 N^2 ||\boldsymbol{G}^{(s,M+R)}( \bar{\boldsymbol{v}}^{(s)})^T  ||^2}.
\end{align}
%
    \begin{figure}[!t]
    \centering
    \includegraphics[width=3.5in,trim={2cm 7cm 1cm 8.5cm},clip]{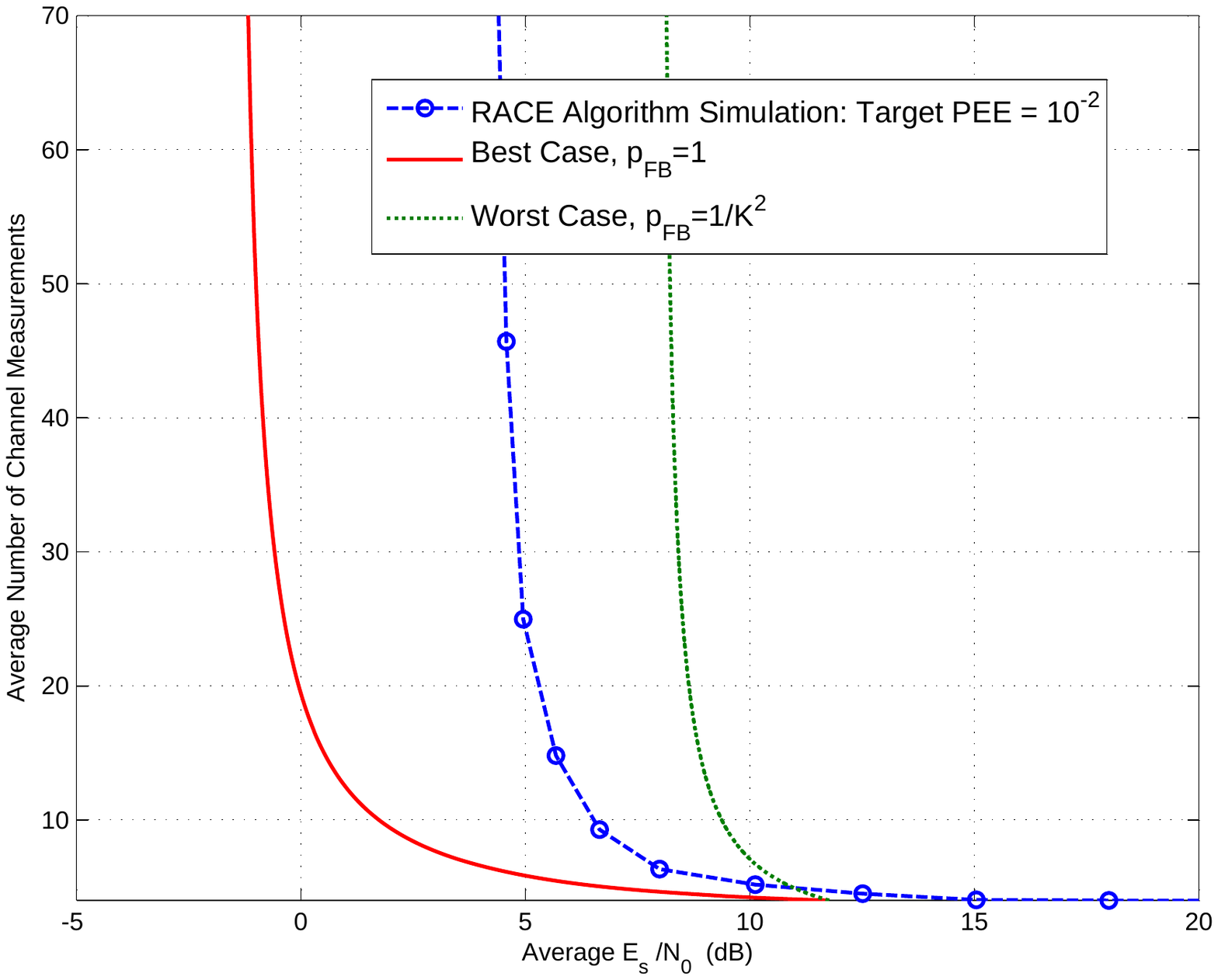}
	\centering
    \caption{Comparison of the minimum number of measurements required for channel estimation in a single stage, using the RACE algorithm, to the analytical expression in (\ref{Rate_4}) with $p_{FB}=1$ and $p_{FB}=1/K^2$. Here we use $N=3$, $K=3$, $||\alpha||=1$ and $M_{max}=\infty$.}
    \label{rate_analyt}
    \end{figure}
The expression in (\ref{Rate_2}) gives a good insight into how to minimize the amount of energy required for successful channel estimation. For example, one intuitive approach to reduce the energy requirement is to maximize the term, $C^4_s N^2|| \boldsymbol{G}^{(s,M+R)}( \bar{\boldsymbol{v}}^{(s)})^T  ||^2$ which directly determines the amount of energy that arrives at the receiver. Furthermore, as we have a minimum number of $M$ initial measurements that have been designed such that they propagate non-zero energy over each sub-range combination (i.e., each column of $\boldsymbol{G}^{(s,M+R)}$ has a non-zero norm), we ensure that $|| \boldsymbol{G}^{(s,M+R)}( \bar{\boldsymbol{v}}^{(s)})^T  ||^2>0, \; \forall \;\bar{\boldsymbol{v}}^{(s)}$. This then guarantees that $E_s/N_0$ always takes on a finite value.

As $\boldsymbol{G}^{(s,M+R)}$ is adaptive and based on feedback, it becomes quite difficult to determine the exact energy requirement for various $R>0$. However, if we let $\boldsymbol{g}^{(s,M+R)}_t$ denote the $t$th row of $\boldsymbol{G}^{(s,M+R)}$, we can expand the term $|| \boldsymbol{G}^{(s,M+R)}( \bar{\boldsymbol{v}}^{(s)})^T  ||^2$ by splitting it into the sum of received energy in each measurement time slots. The row/column energy normalization, seen in the design of (\ref{B_full}), leads to the generator matrix having $||\boldsymbol{G}^{(s,M)}( \bar{\boldsymbol{v}}^{(s)})^T ||^2=\frac{M}{K^2} \; \forall \; \bar{\boldsymbol{v}}^{(s)}$. Expanding this expression we then get
\begin{align}
\label{Rate_simp_3}
&|| \boldsymbol{G}^{(s,M+R)}( \bar{\boldsymbol{v}}^{(s)})^T  ||^2 \nonumber \\&
= \sum \limits_{t=1}^{M+R} \big(\boldsymbol{g}^{(s,M+R)}_t( \bar{\boldsymbol{v}}^{(s)})^T\big)^2 \nonumber\\
&=  ||\boldsymbol{G}^{(s,M)}( \bar{\boldsymbol{v}}^{(s)})^T ||^2+   \sum \limits_{t=M+1}^{M+R} \big(\boldsymbol{g}^{(s,M+R)}_t( \bar{\boldsymbol{v}}^{(s)})^T\big)^2 \nonumber \\
&= \frac{M}{K^2}+   \sum \limits_{t=M+1}^{M+R} \big(\boldsymbol{g}^{(s,M+R)}_t( \bar{\boldsymbol{v}}^{(s)})^T \big)^2.
\end{align}
Following the RACE algorithm, we can also deduce that the term $\boldsymbol{g}^{(s,M+R)}_t( \bar{\boldsymbol{v}}^{(s)})^T$ in (\ref{Rate_simp_3}) results in either unit energy or no energy, depending on whether the most likely sub-range combination from the previous measurement timeslot was correct or not. If we denote the probability of this information being incorrect as $p_{EE|\boldsymbol{G}^{(s,M+R-1)}} $ we can then write the average energy contribution from this term as
\begin{align}
\label{Rate_simp_4}
&|| \boldsymbol{G}^{(s,M+R)}( \bar{\boldsymbol{v}}^{(s)})^T  ||^2  = \frac{M}{K^2}+   \sum \limits_{R'=1}^{R} (1-p_{EE|\boldsymbol{G}^{(s,M+R'-1)}}).
\end{align}
Substituting (\ref{Rate_simp_4}) back into (\ref{Rate_2}), we have
\begin{align}
\label{Rate_3}
 \frac{E_s}{N_0}   \geq \frac{(M+R)^2(2^{\frac{K^2}{M+R}}-1)}{C_s^4 N^2 ||\alpha||^2\Big[\frac{M}{K^2} +   \sum \limits_{R'=1}^{R} (1-p_{EE|\boldsymbol{G}^{(s,M+R'-1)}}) \Big] } .
\end{align}
Unfortunately, the probability in (\ref{Rate_3}) still depends on previous sub-range estimates determined from the previous measurement and is thus difficult to obtain a closed-form. We can however, consider a few extreme cases where we assume a fixed probability of correct feedback information. That is, we use $p_{FB} =1-p_{EE|\boldsymbol{G}^{(s,M+R-1)}}, \; \forall \; R>0$. Finally, we can express the average minimum energy to noise ratio required for successful channel estimation in a given stage with a certain number of measurements to be
\begin{align}
\label{Rate_4}
 \frac{E_s}{N_0}   \geq \frac{(M+R)^2(2^{\frac{K^2}{M+R}}-1)}{C_s^4 N^2 ||\alpha||^2\Big[\frac{M}{K^2} +  R \times p_{FB} \Big] }.
\end{align}
To validate this expression, Fig. \ref{rate_analyt} shows the average number of measurements required to estimate the AOD/AOA information in a single stage for a range of different signal energy to noise ratios. Here we also plot (\ref{Rate_4}) for two special cases. The first is when $p_{FB}=1$, representing the best case scenario where the previous sub-range estimate is always correct. The second is the worst case scenario where the previous sub-range estimate is random (i.e., no feedback) giving $p_{FB}=1/K^2$.

As we can see from Fig. \ref{rate_analyt}, (\ref{Rate_4}) provides a bound for the best case scenario $p_{FB}=1$, and for the worst case scenario $p_{FB}=1/K^2$. To make the simulation results match the bounds, here we do not limit the number of measurements (i.e., $M_{max}=\infty$). Although $p_{FB}=1/K^2$ still gives a mathematical lower bound for the energy requirement, the condition that $p_{FB}=1/K^2$ describes a system that performs far worse than the RACE algorithm. As such, we see the two lines intercept at high SNR. We can also observe from this figure that the RACE algorithm falls between these two curves and converges to $R=0$ at high SNR. Finally, the asymptotic nature of the results in Fig. \ref{rate_analyt} supports the need to set $M_{max} << \infty$ so that measurements do not continue indefinitely.

\subsection{Discussion of System Performance Parameters}

\begin{figure}[!t]
\centering
\includegraphics[width=3.5in,trim={1.5cm 6.5cm 1cm 8.5cm},clip]{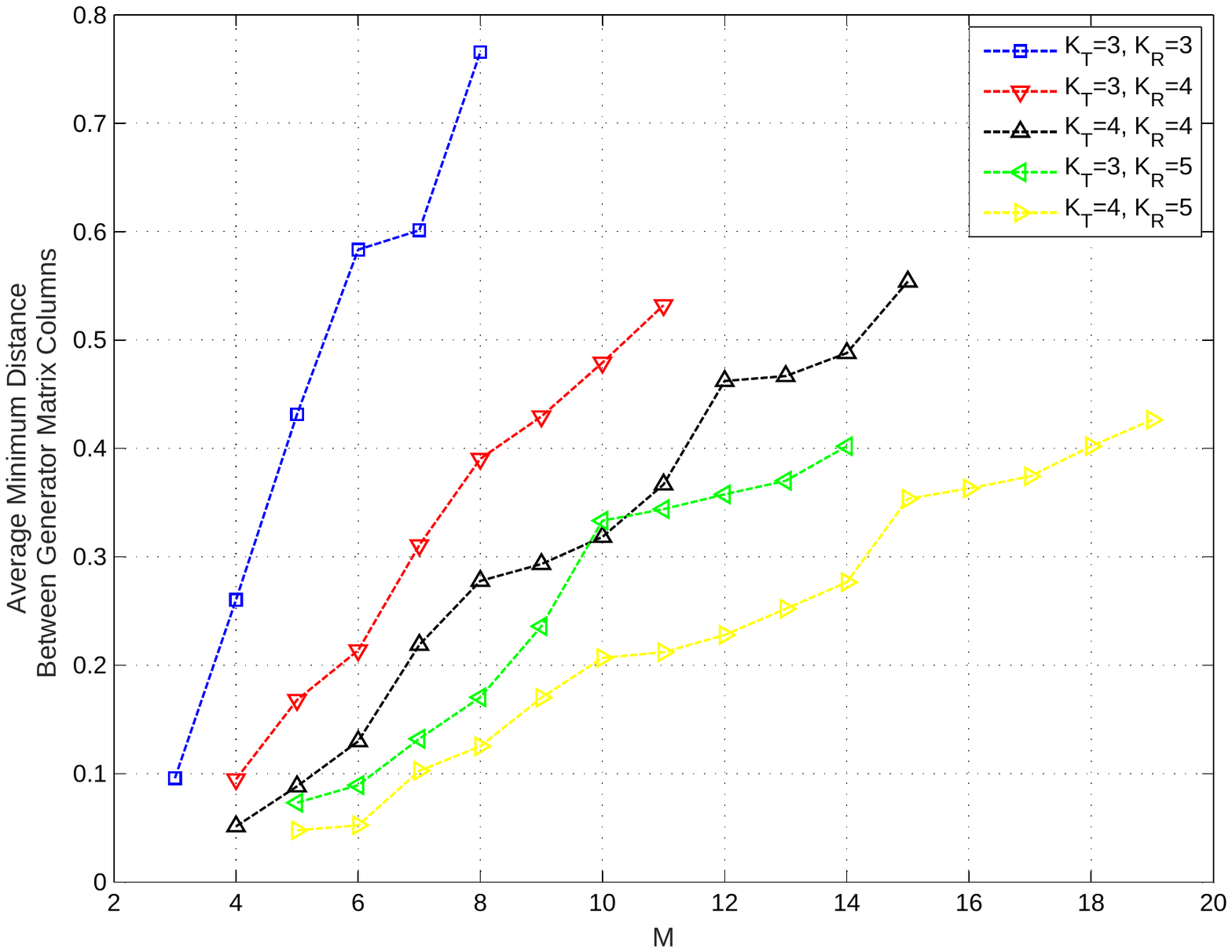}
\caption{ \color{\col} Average minimum distance between the columns of the optimized generator matrix versus the number of measurements, $M$, for a range of different sized beam pattern description matrices.}
\label{g_avg_opt}
\end{figure}

We end this section by briefly discussing the performance parameter selection for the proposed algorithms for a generalized mmWave MIMO system with $N_t$ transmit antenna and $N_r$ receive antenna, each equipped with a limited number of RF chains such that they can only transmit and receive with a single beam pattern.

\subsubsection{Selection of number of sub-ranges in each stage, $K$}
In general, by using a smaller value of $K$, a faster channel estimation can be achieved at a greater energy efficiency and a more computationally efficient ML detection. The main drawback of setting a smaller $K$ lies in the wider beam patterns and the resultant loss of directionality gain. From an energy efficiency point of view, this loss is outweighed by the speed advantage. However, if a peak power constraint were imposed, more directionality may be required in the early stages and this can be achieved by increasing $K$. Selection of $K$ may also depend on the number of antennas being equipped at each link end. For example, it is generally accepted that a mmWave mobile link will have more antennas at a base station (BS) than at the mobile station (MS) \cite{bai2014coverage,hong2014study}. For such a general asymmetric system, it is possible to use different $K$ in each stage and at each transceiver end. This grants flexibility to achieve the same number of estimation stages for each system. Then it is only important that the product of the $K$'s over all stages equals to the antenna number. An alternative approach is to simply use more stages at the BS than at the MS. In this case, once the MS has reached its final stage before the transmitter, it can continue its role in the estimation procedure by utilizing the already estimated AOA.


{\color{\col}
\subsubsection{Selection of $M$ and design of $\boldsymbol{B}_T^{(s,M)}$ and $\boldsymbol{B}_R^{(s,M)}$}

Given that $K$ in each stage has been selected for an arbitrary antenna number $N$, we now propose a method to find the optimal design of the beam pattern description matrices $\boldsymbol{B}_T^{(s,M)}$ and $\boldsymbol{B}_R^{(s,M)}$. To include the asymmetric case where $K$ is different at the transmitter and receiver, we use $K_T$  and $K_R$  to denote the number of sub-ranges at the transmitter and receiver, respectively. We therefore find the optimal design of $\boldsymbol{B}_T^{(s,M)}$ and $\boldsymbol{B}_R^{(s,M)}$ for an arbitrary $K_T$, $K_R$ and $M$ based on the theoretical analysis of PEE conducted in the previous sub-sections. With this result, we then show the performance trade-off for selecting larger/smaller M.

From the analytical expressions (\ref{P_EE})-(\ref{gamma}), we can see that minimizing the PEE is equivalent to maximizing the expression in (\ref{gamma}). We denote the optimal beam pattern description matrices by $\boldsymbol{B}_{T_{opt}}^{(s,M)}$  and $\boldsymbol{B}_{R_{opt}}^{(s,M)}$ at the transmitter and receiver, respectively, and we thus have

\begin{align}
\label{B_opt}
&\{\boldsymbol{B}_{T_{opt}}^{(s,M)},\boldsymbol{B}_{R_{opt}}^{(s,M)}\}= \nonumber \\ &  \;\;\;\;\;\ \underset{{\boldsymbol{B}_T}, \boldsymbol{B}_R}{\operatorname{argmax }}
\sqrt{\frac{P_s C_s^4 N^2 P_R}{ 2 N_0} } || {\boldsymbol{G}^{(s,M)}}(\bar{\boldsymbol{v}}^{(s)}-\bar{\boldsymbol{v}}')^T||, \nonumber \\ \nonumber \\
&\text{s.t., } \text{diag}( \boldsymbol{B}_T (\boldsymbol{B}_T)^H) = \boldsymbol{1},\text{diag}( \boldsymbol{B}_R (\boldsymbol{B}_R)^H) = \boldsymbol{1}, \nonumber \\
&\text{diag}(  (\boldsymbol{B}_T)^H\boldsymbol{B}_T) = \frac{M}{K}\boldsymbol{1},\text{diag}( (\boldsymbol{B}_R)^H \boldsymbol{B}_R ) = \frac{M}{K}\boldsymbol{1}
\end{align}
by recalling that $\boldsymbol{G}^{(s,M)}= \boldsymbol{B}_T^{(s,M)} \odot \boldsymbol{B}_R^{(s,M)}$ and where $\boldsymbol{1}$ is an all-1 vector. The constraints imposed on $\boldsymbol{B}_T^{(s,M)}$  and $\boldsymbol{B}_R^{(s,M)}$  in (\ref{B_opt}-\ref{B_opt_3}), correspond the rows and columns of the matrices having a constant norm. To avoid repetition, we do not list the constraints under (\ref{B_opt_2}-\ref{B_opt_3}), however they still apply. The row and column normalization are imposed, respectively, to:

    \begin{figure*}[!t]
    \centering
    \subfigure[]{\includegraphics[width=3.5in,trim={1.5cm 6.5cm 1cm 8.5cm},clip]{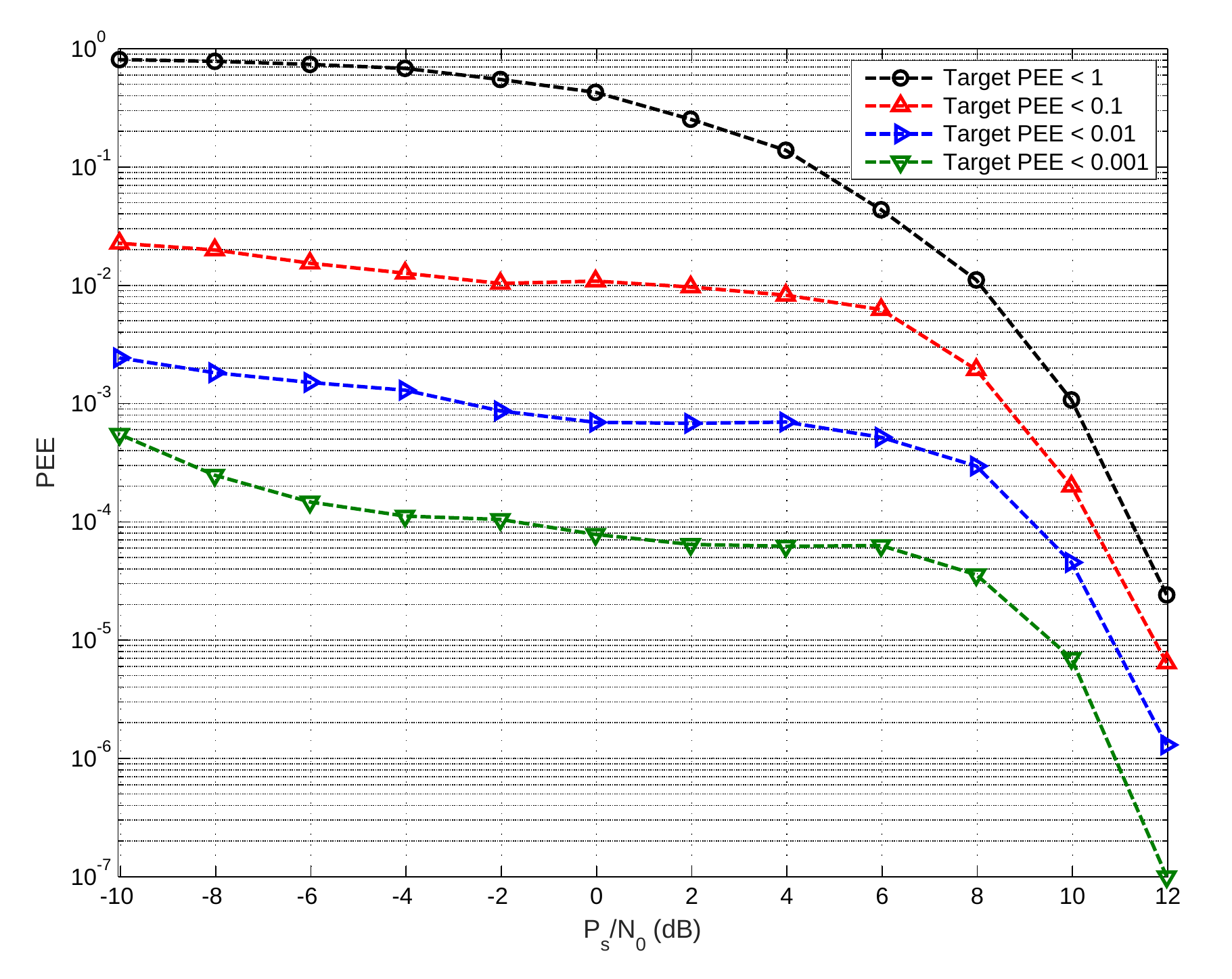}}
    \subfigure[]{\includegraphics[width=3.5in,trim={1.5cm 6.5cm 1cm 8.5cm},clip]{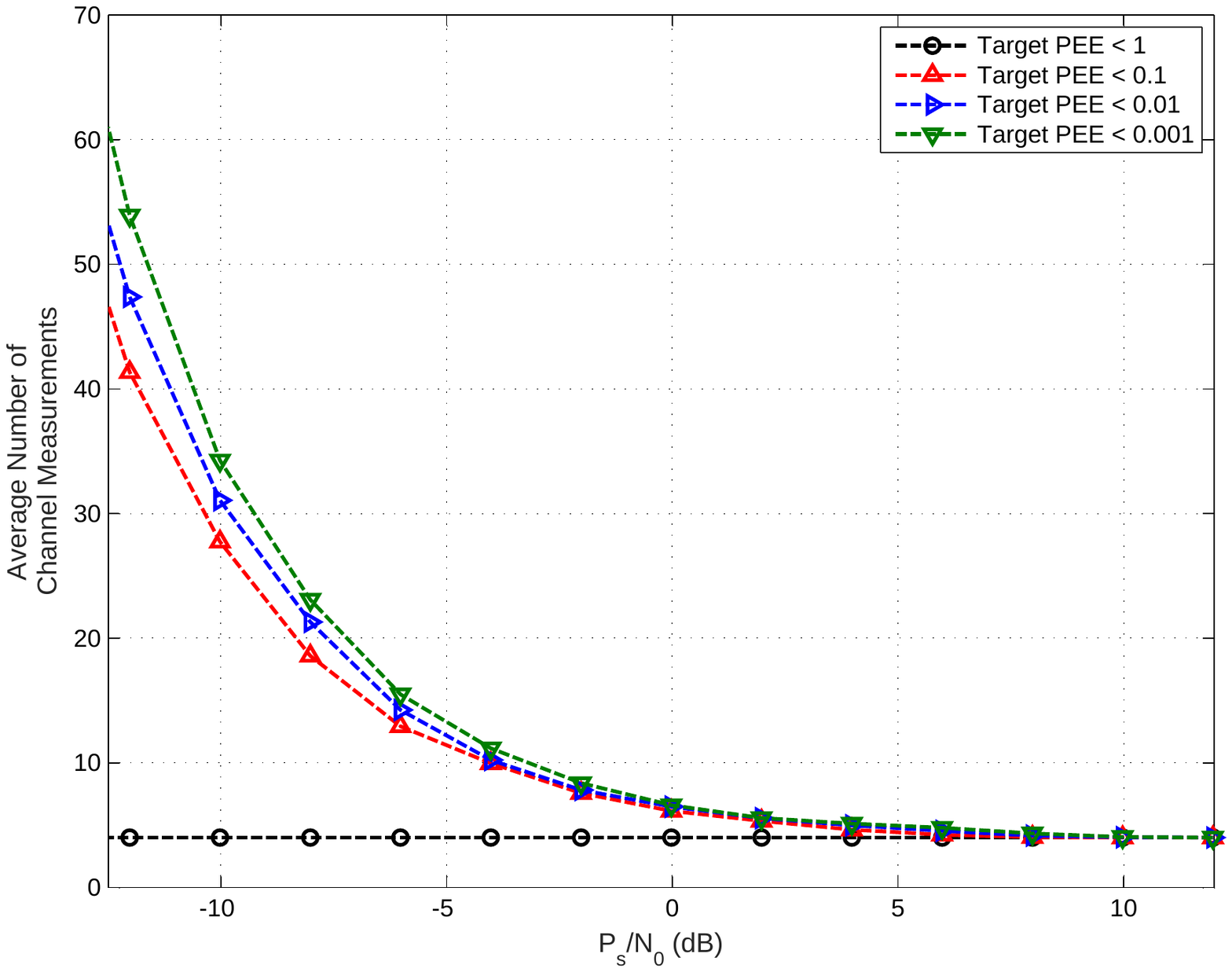}}
    \caption{\color{\col} (a) Probability of estimation error (PEE) results of the RACE algorithm and (b) average number of measurements over a single path AWGN channel with varying target PEE, $\Gamma$}
    \label{AWGN_numerical_res}
    \end{figure*}

\begin{enumerate}
\item Keep a constant transmit power for each measurement of a given stage. It can be achieved when the norms of all rows in $\boldsymbol{B}_T^{(s,M)}$ are the same i.e., $\text{diag}( \boldsymbol{B}_T (\boldsymbol{B}_T)^H) = \boldsymbol{1}$. Similarly, to receive with unit gain, we also have and $\text{diag}( (\boldsymbol{B}_R)^H \boldsymbol{B}_R) = \frac{M}{K}\boldsymbol{1}$.
\item Keep the total energy collected by different AOD/AOA sub-range combinations across all the measurements the same. This ensures the fairness of estimation success in all sub-range combinations. It can be achieved when the norms of all columns in $\boldsymbol{B}_T^{(s,M)}$  and $\boldsymbol{B}_R^{(s,M)}$  are the same, e.g., $\text{diag}(  (\boldsymbol{B}_T)^H)\boldsymbol{B}_T = \frac{M}{K}\boldsymbol{1},\text{diag}( (\boldsymbol{B}_R)^H \boldsymbol{B}_R ) = \frac{M}{K}\boldsymbol{1} $.
\end{enumerate}
Since the square root term is independent of $\boldsymbol{B}_T^{(s,M)}$  and $\boldsymbol{B}_R^{(s,M)}$ for a fixed $K_T$  and $K_R$, we thus can rewrite (\ref{B_opt}) as
\begin{align}
\label{B_opt_2}
\{\boldsymbol{B}_{T_{opt}}^{(s,M)},\boldsymbol{B}_{R_{opt}}^{(s,M)}\}=  \underset{{\boldsymbol{B}_T, \boldsymbol{B}_R}}{\operatorname{argmax }}  || {\boldsymbol{G}^{(s,M)}}(\bar{\boldsymbol{v}}^{(s)}-\bar{\boldsymbol{v}}')^T||.
\end{align}
For channels with a single dominant path, the vectors $\boldsymbol{v}^{(s)}$  and $\boldsymbol{v}'$ only contain a single non-zero element. More specifically, $\boldsymbol{G}^{(s,M)}( \boldsymbol{v}^{(s)}  - \boldsymbol{v}' )$ can be simplified to the subtraction of two columns of $\boldsymbol{G}^{(s,M)}$. Since the channel estimation error is dominated by the columns of $\boldsymbol{G}^{(s,M)}$ with the smallest Euclidean distance, we can then re-write (\ref{B_opt_2}) as,
\begin{align}
\label{B_opt_3}
\{\boldsymbol{B}_{T_{opt}}^{(s,M)},\boldsymbol{B}_{R_{opt}}^{(s,M)}\} &= \nonumber \\   \underset{{\boldsymbol{B}_T, \boldsymbol{B}_R}}{\operatorname{argmax }}  & \Big( \underset{ \begin{array}{c} i,j \\
i \neq  j \end{array} }{\operatorname{min }} || {\boldsymbol{g}_i^{(s,M)}} -{\boldsymbol{g}_j^{(s,M)}} || \Big).
\end{align}
by recalling that $\boldsymbol{g}_i^{(s,M)}$ represents the $i$th column of the matrix $\boldsymbol{G}^{(s,M)}$. In order to find a solution for $\boldsymbol{B}_{T_{opt}}^{(s,M)}$, $\boldsymbol{B}_{R_{opt}}^{(s,M)}$, we then define $\mathcal{B}_b^{(M,K)}=\{ \boldsymbol{B}_b^{(1) },\boldsymbol{B}_b^{(2) },\cdots \}$ as the set of all possible $M \times K$ binarized beam pattern description matrices. In order to approximate the normalization constraints, we first normalize the columns of all matrices in this set followed by the normalization of the rows. We denote this normalized set by $\mathcal{B}^{(M,K)}=\{ \boldsymbol{B}^{(1) },\boldsymbol{B}^{(2) },\cdots \}$. We can then carry out an offline search within this set to achieve the best solution $\boldsymbol{B}_{T_{opt}}^{(s,M)}$ and $\boldsymbol{B}_{R_{opt}}^{(s,M)}$ i.e., by comparing all combinatorial pairs of $\boldsymbol{B}_T^{(s,M)} \in  \mathcal{B}^{(M,K_T)}, \boldsymbol{B}_R^{(s,M)} \in  \mathcal{B}^{(M,K_R)}$. It should be noted that there might be multiple solutions to the optimal beam pattern description matrices $\boldsymbol{B}_{T_{opt}}^{(s,M)}$ and $\boldsymbol{B}_{R_{opt}}^{(s,M)}$, which simply have different column permutations but result in the same performance.

For larger $M$ and $K$, the complexity of the above search-based approach can be reduced by further constraining the set $\mathcal{B}_b^{(M,K)}$. For example, to keep the directivity gains similar in all beam patterns in each stage, we fix the number of non-zero sub-ranges at the transmitter and receiver to $W_T$ and $W_R$, respectively. To impose this constraint, we reduce $\mathcal{B}_b^{(M,K)}$ to the set of all $M \times K$ binarized beam patterns that have row weights of $W_T$ at the transmitter and $W_R$ at the receiver. Here, row weight refers to the number of non-zero entries in each row of a matrix. For the example beam patterns given in (\ref{G_K_3_2}), we have a row weight of $W_R=W_T=2$. In general, transmitting/measuring on a greater number of sub-ranges in each measurement will lead to a lower minimum number of measurements required for estimation. This is because all sub-range combinations can be spanned in a fewer number measurements. This choice, however, will lead to less directional beam patterns, and therefore a loss of directivity gain.

    \begin{figure*}[!t]
    \centering
    \subfigure[]{\includegraphics[width=3.5in,trim={1.5cm 6.5cm 1cm 8.5cm},clip]{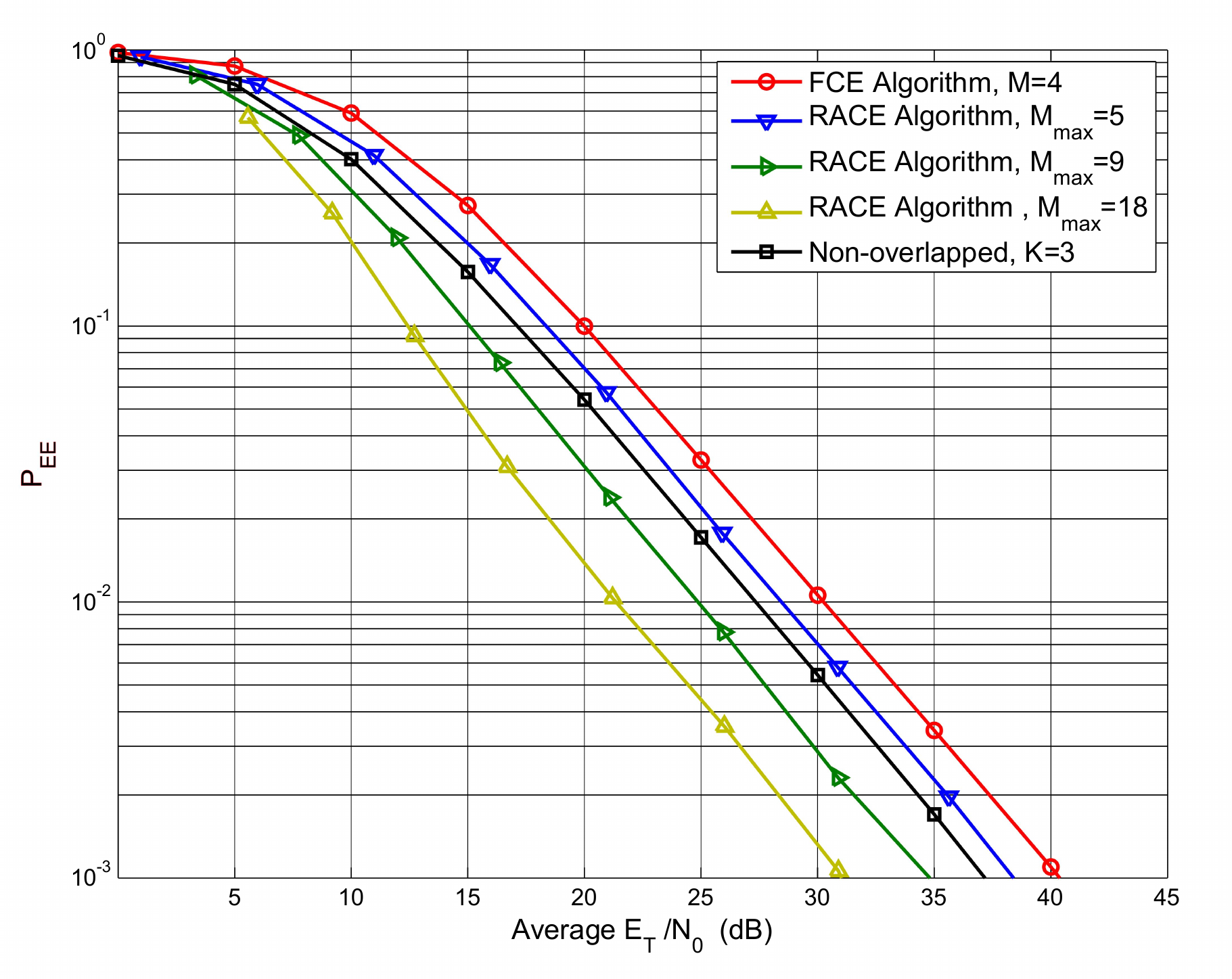}}
    \subfigure[]{\includegraphics[width=3.5in,trim={1.5cm 6.5cm 1cm 8.5cm},clip]{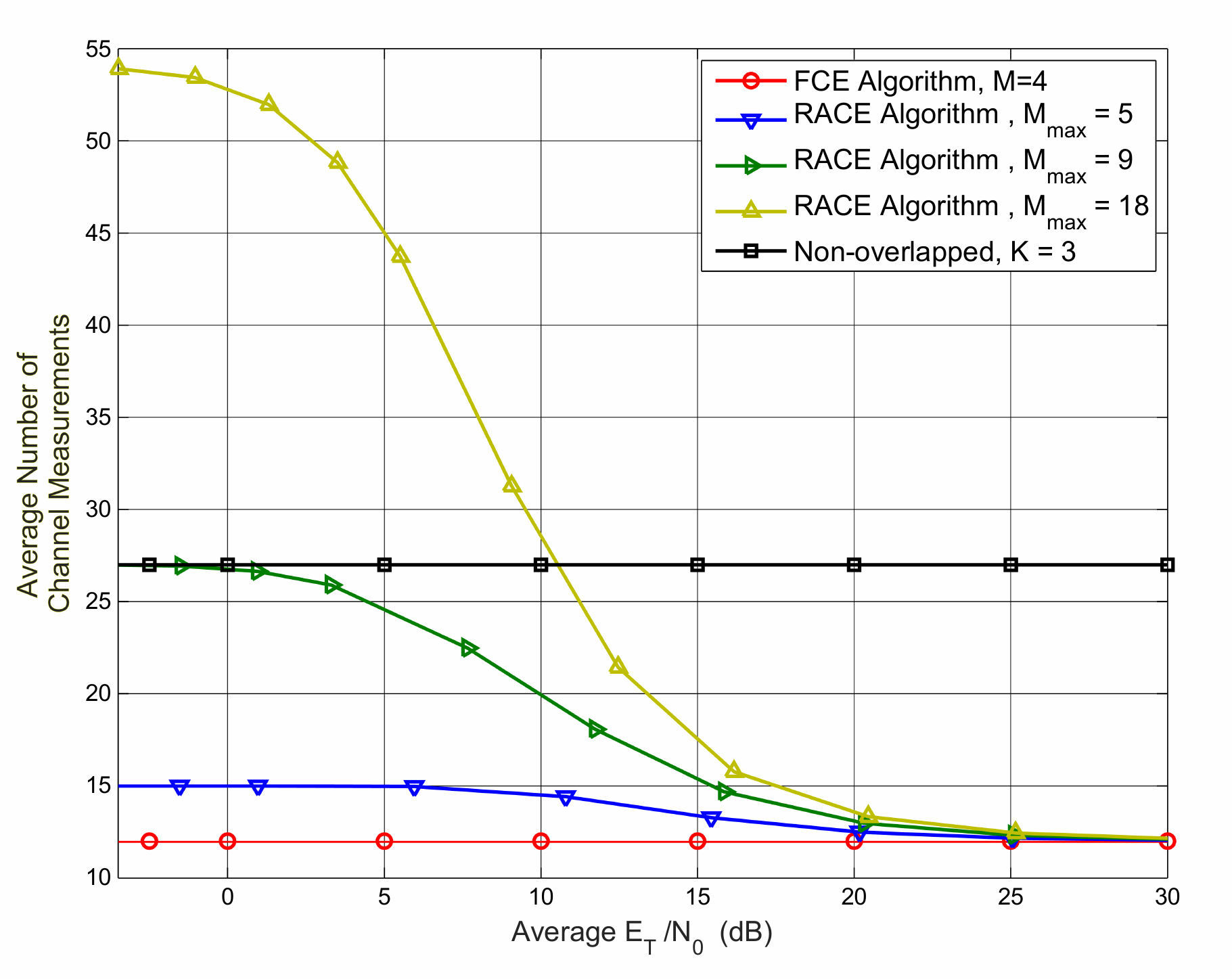}}
    \caption{Numerical performance results of the proposed algorithms compared to the non-overlapped algorithm presented in \cite{rheath} for (a) probability of estimation error (PEE) and (b) average number of measurements required for channel estimation.}
    \label{numerical_res}
    \end{figure*}

By using the aforementioned approach and by denoting $K_T$ and $K_R$  as the number of sub-ranges at the transmitter and receiver respectively, Fig. \ref{g_avg_opt} shows the average minimum Euclidean distance between the columns of the generator matrix versus $M$ for a range of different $K_T$ and $K_R$. Here we constrain the row weights at the transmitter to be $W_T=K_T-1$ and that at the receiver to be $W_R=K_R-1$. We can see from this figure that for a fixed $K_T$ and $K_R$, adding additional measurements can increase the spatial separation of the columns in the generator matrix. Based on our analysis, this in turn decreases the probability that one path is mistakenly estimated as another, therefore improving the overall estimation performance. On the other hand, using a larger $M$ will require a greater number of channel measurements in each stage. If the FCE algorithm is used, we can increase the value of $M$ to improve the estimation performance. If the RACE algorithm is employed, it is generally desired to set $M$ relatively low, resulting in a fast estimation that is then confirmed, if needed, by additional measurements. It can also be seen that, when larger values of $K_T$  and $K_R$  are employed with a similar $M$, the average minimum Euclidean distances are found to be smaller. This is because these parameters correspond to the estimation of a larger number of sub-range combinations in each stage. Although the energy required for estimation in a single stage may need to be increased for a fixed PEE, the number of over all stages would be less, due to the increased values of $K$}.


\subsubsection{Selection of $M_{max}$ and $\Gamma$  }

Finally, the selection of the maximum number of measurements per stage $M_{max}$ should be selected based upon any maximum timing constraints. Increasing $M_{max}$ will significantly increase the energy efficiency at medium to high SNR regime, although the average number of measurements per stage will still converge to $M$ at high SNR. We find that, for a fading channel with Gaussian distributed channel coefficients, the selection of $\Gamma$ is not as important as $M_{max}$. This is due to the non-zero probability of an outage condition. In general, increasing $\Gamma$ will reduce the PEE by increasing the \textit{average} number of measurements and therefore expending more energy. In contrast, at high SNR, increasing $M_{max}$ will allow more measurements to be carried out in the unlikely event that it needs them. Therefore the later uses less average energy.

\section{Numerical Results}

We now provide some numerical examples to verify the performance of our proposed algorithms. {{\color{\col}First, we evaluate the impact of the target PEE, $\Gamma$, on channel estimation performance by comparing the results of the RACE algorithm over a simple single path AWGN channel with $K=3$, $M=4$, and $N=3$ (i.e., single stage). Here we use $|\alpha|=P_R=1$ with varying $\Gamma$.  To show the constant average PEE, here we set $M_{max}  =250$ so that even at low SNR, the achieved PEE is unaffected by the number of measurements saturating to its maximum. In addition, as the channel is AWGN, there will also be no chance of a ``deep fade” where measurements may continue indefinitely. From Fig. \ref{AWGN_numerical_res} (a) and (b), we can see that the RACE algorithm increases the number of measurements in attempt to hold the PEE below the target value. We also see that the PEE achieved is normally lower than the target PEE. This is largely because the target PEE is actually the highest allowable PEE, for which no additional measurements are required. As such, the average PEE is the average of many channel estimations with performance better than this target. Furthermore, in some cases (particularly in mid-SNR where the number of measurements are low), a single additional measurement on the correct sub-range will significantly increase the PEE beyond the target PEE. We see that this effect diminishes at low SNR where the significance of additional measurements is not as large as that in medium to high SNR. That is, the average PEE is closer to the target PEE.}

We now consider a mmWave system with $N = 27$ antennas at both the transmitter and receiver. We use a single path channel with fading coefficient, $\alpha$, assumed to follow a complex Gaussian distribution with zero mean and variance $P_R=1$. We assume the corresponding AOD, $\phi_l^t \in \mathcal{U}_N $, and AOA, $\phi_l^t \in \mathcal{U}_N $, to follow a random uniform distribution. We set $K = 3$ for both of our algorithms and the non-overlapped multi-stage algorithm in \cite{rheath}, all requiring $S=3$ stages. $K^2 = 9$ measurements are required in each stage of \cite{rheath}. However, in our proposed algorithms, only $M = 4$ time slots are involved in each stage the FCE algorithm and a minimum of $M = 4$ for the RACE algorithm. The performance of the RACE algorithm is shown for a number of different values of maximum measurements, $M_{max}$, and uses a target PEE of $\Gamma=10^{-2}$. Power allocation among $S=3$ stages is applied to all algorithms as (\ref{p_s}).

Fig. \ref{numerical_res}(a) shows the probability of an incorrect AOD/AOA estimate after the $S$ stages of estimation have been carried out and Fig. \ref{numerical_res}(b) shows the average total number of measurements required for the same estimation. The total energy required in the overall channel estimation process is calculated by $E_T=\sum_{s=1}^S P_s (M+R)$. We can see that, to achieve the same PEE as that in \cite{rheath}, the FCE algorithm requires 2.5dB more energy for a given noise power $N_0$. However, the number of required channel measurements is decreased by a factor of 2.25.

For the RACE algorithm, we can also see that with $M_{max}=5$, while still significantly faster than \cite{rheath}, it has an improved energy gain for a given PEE compared to $M_{max}=4$, but still worse than than  \cite{rheath} by about 1dB.  When $M_{max}$ is increased to $M_{max}=K^2=9$, we can observe from Fig. \ref{numerical_res}(a) that the required energy in this case is around 2.5dB better than \cite{rheath}. From Fig. \ref{numerical_res}(b) we see that when $M_{max}=K^2=9$ is used, the RACE algorithm is always faster than \cite{rheath} and, at medium to high SNR (i.e., $E_T/N_0>25$), the average number of channel measurements converges to $SM$. That is, using the proposed RACE algorithm with $M_{max}=K^2$ achieves 2.25 factor reduction of channel measurements while also achieving an energy gain of $2.5$dB for a given probability of error. Further increasing the maximum number of measurements to $M_{max}=2K^2=18$ (i.e., the algorithm at most requires twice the measurements required of \cite{rheath}) we see that the RACE algorithm achieves an energy gain of 6dB compared to \cite{rheath} at medium to high SNR, while requiring an average of 2.25 times less channel measurements. Furthermore, we can observe from SNR's greater than 10.5dB that the average number of measurements required in the RACE algorithm is also less than that required by the algorithm in \cite{rheath}.
    \begin{figure}[!t]
    \centering
    \includegraphics[width=3.45in,trim={2cm 6.5cm 1cm 8.5cm},clip]{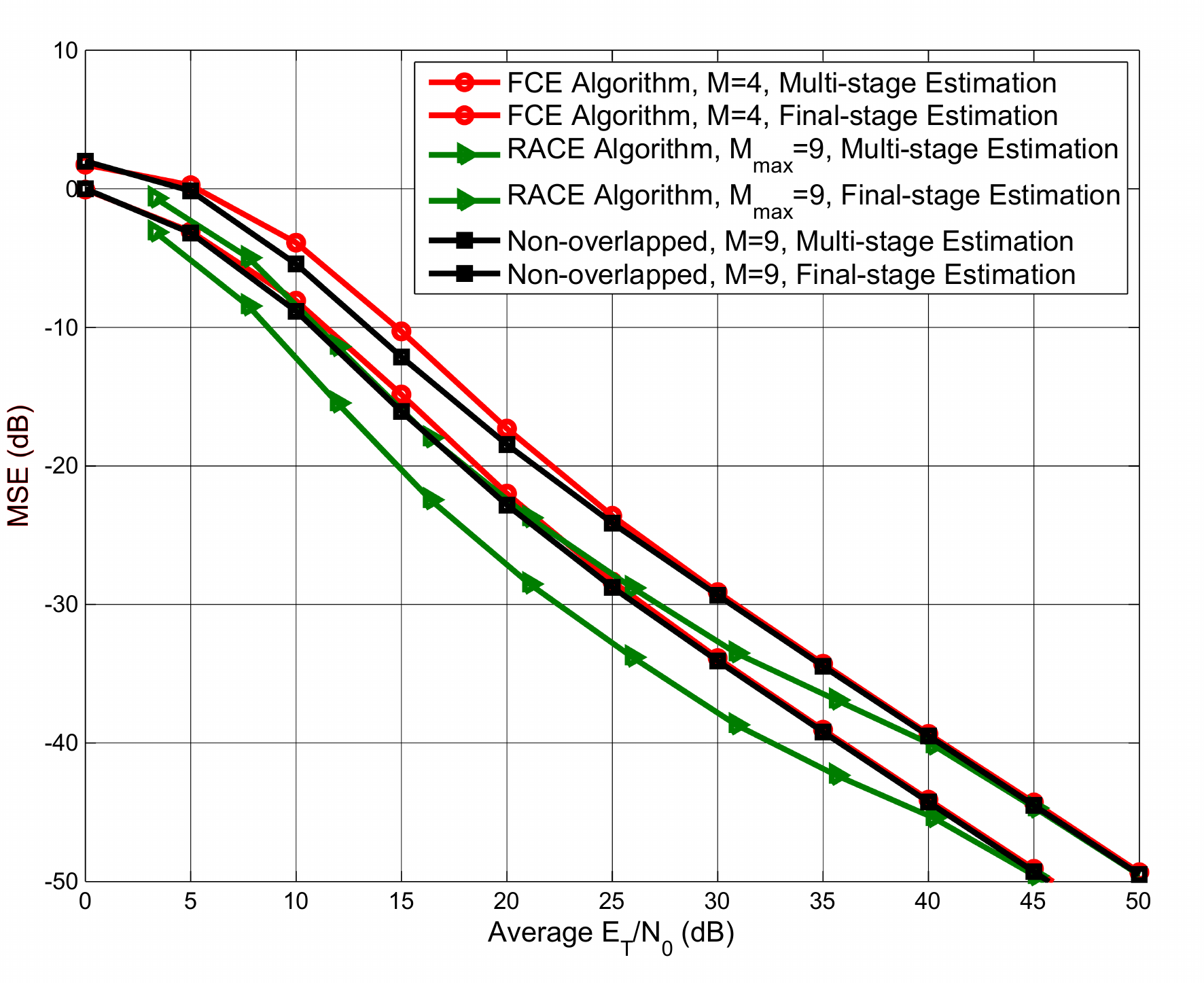}
	\centering
    \caption{Numerical performance results of the proposed algorithms compared to the non-overlapped presented in \cite{rheath} for estimation of the fading coefficient, $\alpha$. Results are shown for estimation using just the final stage of estimation and for the the multi-stage approach shown in  \ref{sssec:MMSE}. The average estimation error is expressed in dB as $\text{E}[{|\alpha- \hat{\alpha} |^2]}$.}
    \label{MSE_fig}
    \end{figure}

Fig. \ref{MSE_fig} shows the relative MSE in dB of the fading coefficient $\alpha$, i.e., $\text{E}[{| \alpha-\hat{\alpha} |^2]}$. As can be seen, the LMMSE estimator proposed in section \ref{sssec:MMSE} is compared to the one only using measurements from final stage estimation. We can see that the multi-stage LMMSE estimator improves the performance accuracy of the fading coefficient estimation for all algorithms. {\color{\col}In the low SNR range, we also see that the proposed FCE algorithm has slightly worse estimation accuracy when compared to \cite{rheath}. This is predominantly caused by the FCE algorithm having a worse PEE for the same SNR. This performance loss diminishes at mid to high SNR due to the inherent spread of energy over multiple sub-range combinations in the overlapped beam pattern design. For example, in the event that the FCE algorithm selects an incorrect angular range in the final stage, the next most likely sub-range will usually still contain some information of $\alpha$. Focusing on the medium SNR range, Fig. \ref{MSE_fig} also shows that the RACE algorithm with $M_{max}=K^2=9$ is able to estimate the fading coefficient more accurately when compared to both the FCE algorithm and the algorithm in \cite{rheath}. This is because, after  the additional measurements are directed onto the most likely propagation path. This, in turn, increases the final estimation accuracy once the AOD/AOA have been determined. As the average number of these additional measurements converges to zero at high SNR, this performance gain is lost at high SNR. Interestingly, at high SNR, it can be seen that all approaches converge to a similar path coefficient MSE performance, despite having different PEE. This is largely because when each algorithm identifies an incorrect AOA/AOA, it is usually when the channel is in a deep fade. As a result, misestimating the path coefficient in this case has little effect on overall MSE of the channel fading coefficient. As such, PEE is really a better performance metric for beam misalignment. }

{\color{\col}We now test the performance of the proposed algorithms under multipath scenarios. Similarly, with \cite{rheath}, we simulate the cases with $L=2$ paths, shown in Fig. \ref{L_2_sim} (a) and (b), respectively and with $L=3$ paths, shown in Fig. \ref{L_3_sim} (a) and (b), respectively. In these figures, the most immediate observation is that the FCE algorithm reaches an error floor at high SNR. This is mainly because the non-zero probability of two or more paths having similar magnitude and being in both the non-overlapped sub-ranges. In this scenario, the received measurement vector will be similar to one very strong path in the overlapped region. This is largely the motivation for the RACE algorithm as it permits an additional measurement to confirm the true sub-range combination corresponding to one of the propagation paths. As can be observed in these two figures, the RACE algorithm is able to seamlessly adapt to this scenario, still yielding up to a 5dB gain compared to \cite{rheath} and converging on being almost 2.25 times faster on average. It is worth noting that this time advantage is only possible with the initial overlapped beam pattern design of the FCE algorithm. That is, the RACE algorithm still needs the initial beam pattern design of FCE to be faster than \cite{rheath}. Finally, similarly to \cite{rheath}, we can also see that the performance of RACE increases slightly as the number of paths increases from $L=1$ to $L=3$. This is understandable since the chance of detecting one of multiple paths is greater than the chance of detecting a single one.}

\section{Conclusion}

In this paper we have proposed a novel fast channel estimation algorithm for mmWave communication systems based on a novel overlapped beam pattern design. In general, our proposed algorithm can speed up the channel estimation process by a factor of $K^2/M$ when compared to existing algorithms with non-overlapped beam patterns. Using a fixed number of measurements, we show that this reduction of measurements comes at an energy-to-noise ratio expense of 2.5dB in order to achieve the same probability of estimation error (PEE). For channels with rapidly changing channel information, this cost may be justified in order to improve the estimation speed. We have also proposed a novel rate-adaptive channel estimation algorithm, in which additional measurements are carried out when a high probability of estimation error is expected. We show that by taking this approach, the channel can be estimated more efficiently, yielding significant gains of up to 6dB when compared to the algorithm in \cite{rheath}, while still converging to the same average number of measurements as the fast channel estimation (FCE) algorithm at high SNR.

    \begin{figure*}[!t]
    \centering
    \subfigure[]{\includegraphics[width=3.5in,trim={1.5cm 6.5cm 1cm 8.5cm},clip]{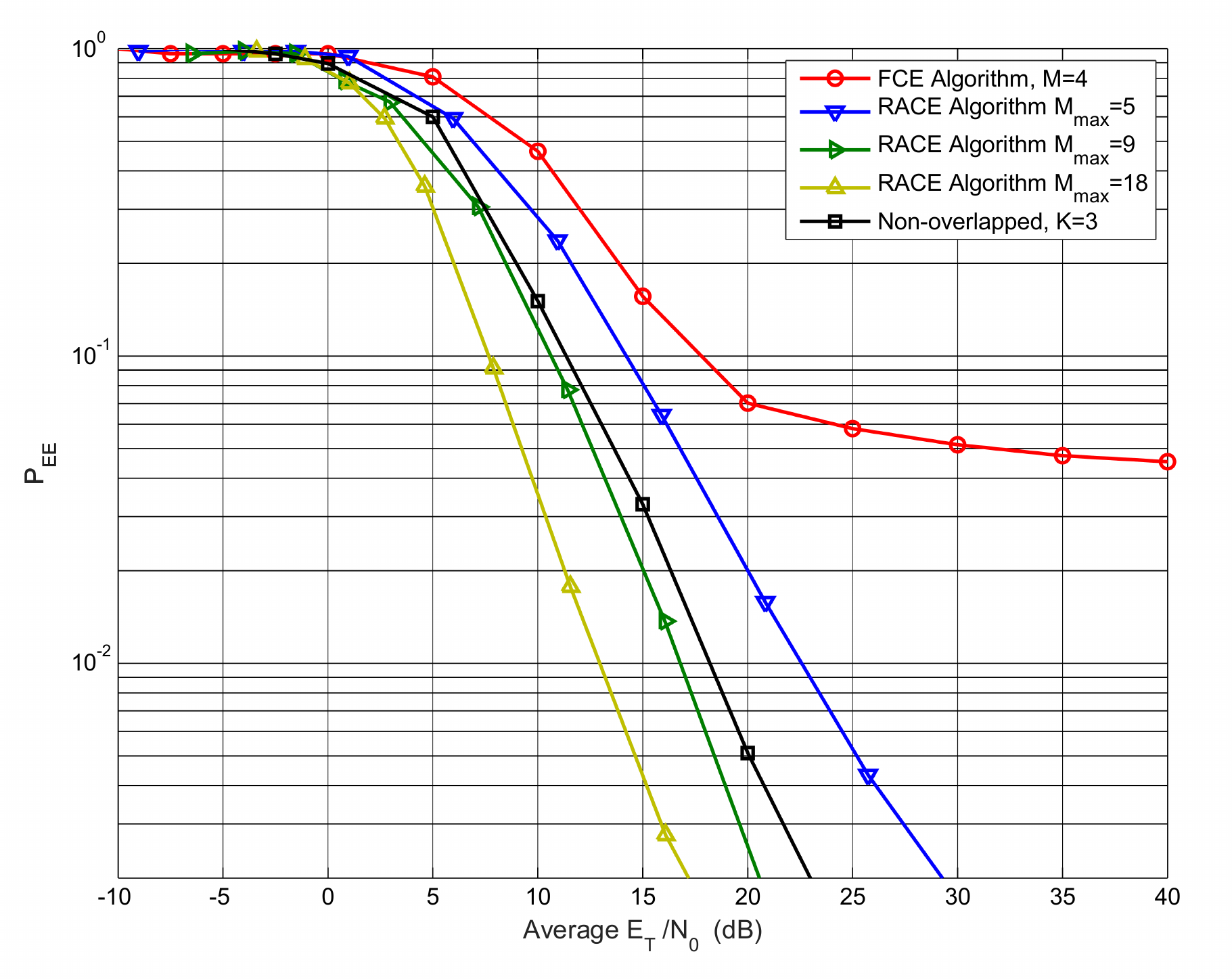}}
    \subfigure[]{\includegraphics[width=3.5in,trim={1.5cm 6.5cm 1cm 8.5cm},clip]{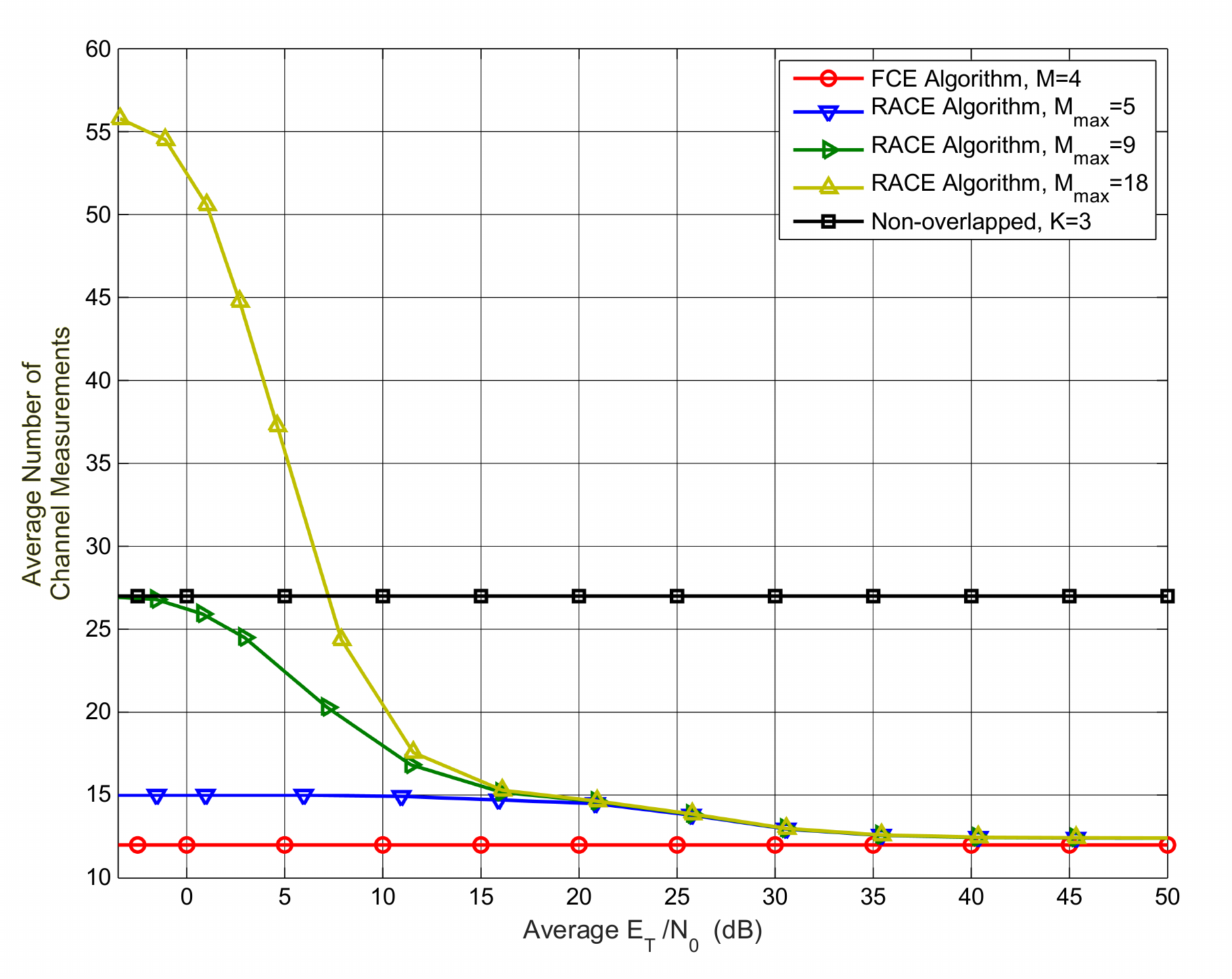}}
    \caption{{\color{\col}The performance comparison between the proposed algorithms with those in \cite{rheath} in terms of (a) probability of estimation error (PEE) and (b) average number of measurements required for channel estimation for the case $L=2$.}}
    \label{L_2_sim}
    \end{figure*}

    \begin{figure*}[!t]
    \centering
    \subfigure[]{\includegraphics[width=3.5in,trim={1.5cm 6.5cm 1cm 8.5cm},clip]{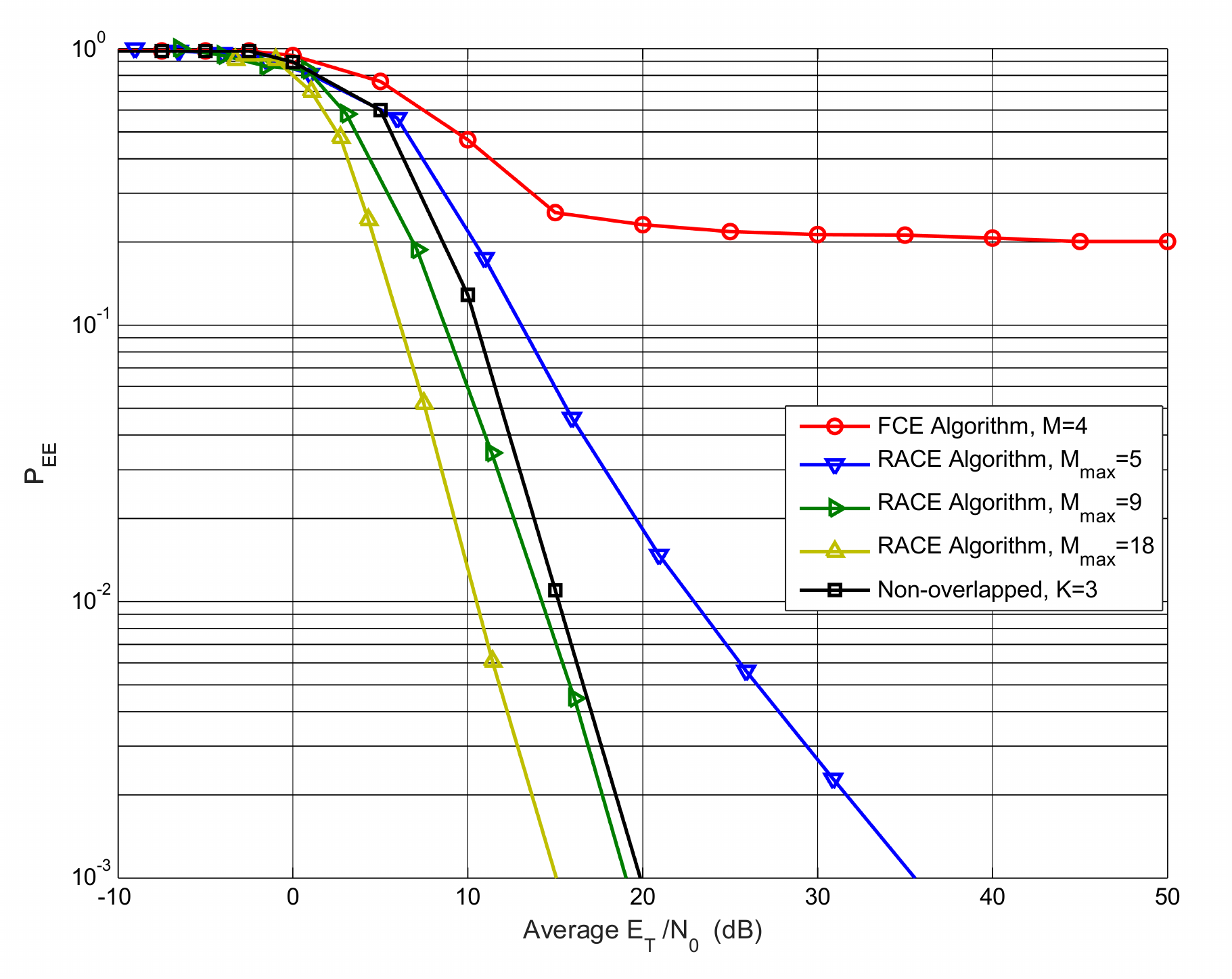}}
    \subfigure[]{\includegraphics[width=3.5in,trim={1.5cm 6.5cm 1cm 8.5cm},clip]{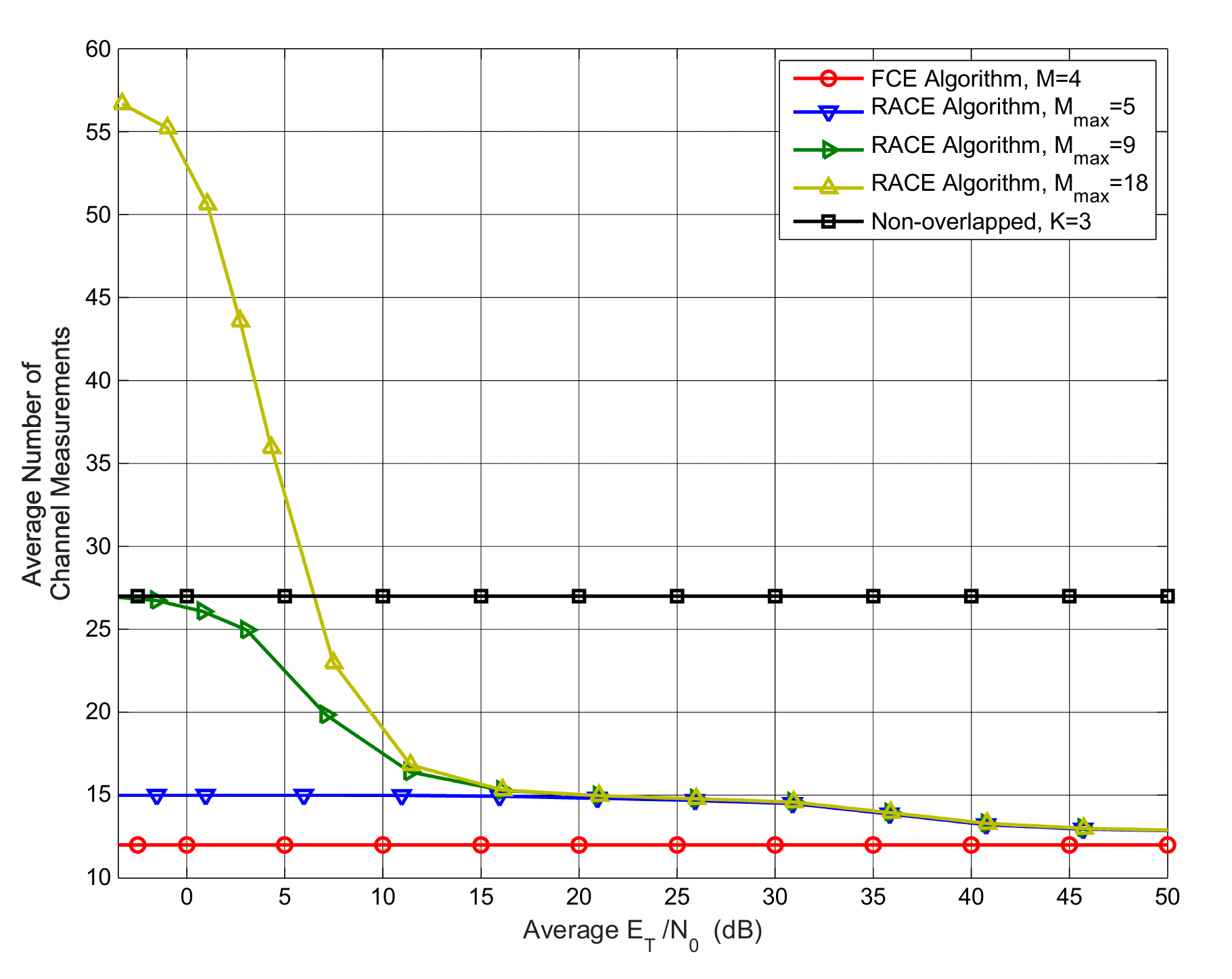}}
    \caption{{\color{\col}The performance comparison between the proposed algorithms with those in \cite{rheath} in terms of (a) probability of estimation error (PEE) and (b) average number of measurements required for channel estimation for the case $L=3$.}}
    \label{L_3_sim}
    \end{figure*}

\bibliographystyle{IEEEtran}
\bibliography{IEEEabrv,TSP}

\begin{thebibliography}{10}
\providecommand{\url}[1]{#1}
\csname url@samestyle\endcsname
\providecommand{\newblock}{\relax}
\providecommand{\bibinfo}[2]{#2}
\providecommand{\BIBentrySTDinterwordspacing}{\spaceskip=0pt\relax}
\providecommand{\BIBentryALTinterwordstretchfactor}{4}
\providecommand{\BIBentryALTinterwordspacing}{\spaceskip=\fontdimen2\font plus
\BIBentryALTinterwordstretchfactor\fontdimen3\font minus
  \fontdimen4\font\relax}
\providecommand{\BIBforeignlanguage}[2]{{%
\expandafter\ifx\csname l@#1\endcsname\relax
\typeout{** WARNING: IEEEtran.bst: No hyphenation pattern has been}%
\typeout{** loaded for the language `#1'. Using the pattern for}%
\typeout{** the default language instead.}%
\else
\language=\csname l@#1\endcsname
\fi
#2}}
\providecommand{\BIBdecl}{\relax}
\BIBdecl

\bibitem{Kokshoorn}
M.~Kokshoorn, P.~Wang, Y.~Li, and B.~Vucetic, ``Fast channel estimation for
  millimetre wave wireless systems using overlapped beam patterns,'' in
  \emph{IEEE Int. Conf. on Commun. (ICC)}, June 2015, pp. 1304--1309.

\bibitem{Koks1612}
M.~Kokshoorn, H.~Chen, Y.~Li, and B.~Vucetic, ``{RACE:} a rate adaptive channel
  estimation approach for millimeter wave {MIMO} systems,'' in \emph{Proc. IEEE
  Global Telecomm. Conf.}\hskip 1em plus 0.5em minus 0.4em\relax IEEE, Dec.
  2016, pp. 1--6.

\bibitem{pi2011introduction}
Z.~Pi and F.~Khan, ``An introduction to millimeter-wave mobile broadband
  systems,'' \emph{IEEE Commun. Mag.}, vol.~49, no.~6, pp. 101--107, June 2011.

\bibitem{rappaport2013millimeter}
T.~S. Rappaport, S.~Sun, R.~Mayzus, H.~Zhao, Y.~Azar, K.~Wang, G.~N. Wong,
  J.~K. Schulz, M.~Samimi, and F.~Gutierrez, ``Millimeter wave mobile
  communications for 5{G} cellular: It will work!'' \emph{IEEE Access}, vol.~1,
  pp. 335--349, May 2013.

\bibitem{rheath}
A.~Alkhateeb, O.~El~Ayach, G.~Leus, and R.~Heath, ``Channel estimation and
  hybrid precoding for millimeter wave cellular systems,'' \emph{IEEE J. Sel.
  Topics Signal Process.}, vol.~8, no.~5, pp. 831--846, Oct. 2014.

\bibitem{hong2014study}
W.~Hong, K.-H. Baek, Y.~Lee, Y.~Kim, and S.-T. Ko, ``Study and prototyping of
  practically large-scale mmwave antenna systems for 5g cellular devices,''
  \emph{IEEE Commun. Mag.}, vol.~52, no.~9, pp. 63--69, 2014.

\bibitem{Rappaport200}
S.~Rangan, T.~Rappaport, and E.~Erkip, ``Millimeter-wave cellular wireless
  networks: Potentials and challenges,'' \emph{Proc. IEEE}, vol. 102, no.~3,
  pp. 366--385, March 2014.

\bibitem{zhang2010channel}
H.~Zhang, S.~Venkateswaran, and U.~Madhow, ``Channel modeling and {MIMO}
  capacity for outdoor millimeter wave links,'' in \emph{Proc. Wireless Commun.
  Netw. Conf. (WCNC)}, Apr. 2010, pp. 1--6.

\bibitem{torkildson2010channel}
E.~Torkildson, H.~Zhang, and U.~Madhow, ``Channel modeling for millimeter wave
  mimo,'' in \emph{Proc. Inf. Theory Appl. Workshop}.\hskip 1em plus 0.5em
  minus 0.4em\relax IEEE, 2010, pp. 1--8.

\bibitem{hur2013millimeter}
S.~Hur, T.~Kim, D.~Love, J.~Krogmeier, T.~Thomas, and A.~Ghosh, ``Millimeter
  wave beamforming for wireless backhaul and access in small cell networks,''
  \emph{IEEE Trans. Commun.}, vol.~61, no.~10, pp. 4391--4403, Oct. 2013.

\bibitem{biglarbegian2011optimized}
B.~Biglarbegian, M.~Fakharzadeh, D.~Busuioc, M.-R. Nezhad-Ahmadi, and
  S.~Safavi-Naeini, ``Optimized microstrip antenna arrays for emerging
  millimeter-wave wireless applications,'' \emph{IEEE Trans. Antennas Propag.},
  vol.~59, no.~5, pp. 1742--1747, 2011.

\bibitem{5284444}
``Wireless medium access control ({MAC}) and physical layer ({PHY})
  specifications for high rate wireless personal area networks ({WPANs}),
  amendement 2: Millimeter-wave-based alternative physical layer extension,''
  \emph{IEEE Std 802.15.3c}, pp. c1--187, Oct 2009.

\bibitem{beamcodebook}
J.~Wang, Z.~Lan, C.-W. Pyo, T.~Baykas, C.-S. Sum, M.~Azizur~Rahman, R.~Funada,
  F.~Kojima, I.~Lakkis, H.~Harada, and S.~Kato, ``Beam codebook based
  beamforming protocol for multi-gbps millimeter-wave {WPAN} systems,'' in
  \emph{IEEE J. Select. Areas Commun.}, 2009, pp. 1390--1399.

\bibitem{chen2011multi}
L.~Chen, Y.~Yang, X.~Chen, and W.~Wang, ``Multi-stage beamforming codebook for
  60{GH}z {WPAN},'' in \emph{Proc. 6th Int. ICST Conf. Commun. Network. China},
  Aug. 2011, pp. 361--365.

\bibitem{sayeed2002deconstructing}
A.~Sayeed, ``Deconstructing multiantenna fading channels,'' \emph{IEEE Trans.
  Sig. Process.}, vol.~50, no.~10, pp. 2563--2579, Oct. 2002.

\bibitem{hong2003spatial}
Z.~Hong, K.~Liu, R.~W. Heath~Jr, and A.~M. Sayeed, ``Spatial multiplexing in
  correlated fading via the virtual channel representation,'' \emph{IEEE J.
  Select. Areas Commun.,}, vol.~21, no.~5, pp. 856--866, 2003.

\bibitem{lagarias1998convergence}
J.~C. Lagarias, J.~A. Reeds, M.~H. Wright, and P.~E. Wright, ``Convergence
  properties of the nelder--mead simplex method in low dimensions,'' \emph{SIAM
  J. Optim.}, vol.~9, no.~1, pp. 112--147, 1998.

\bibitem{Jianhua}
J.~Mo, P.~Schniter, N.~Gonzalez~Prelcic, and R.~Heath, ``Channel estimation in
  millimeter wave mimo systems with one-bit quantization,'' in \emph{Proc.
  Conf. Signals, Syst. Comput.}, Nov 2014, pp. 957--961.

\bibitem{Seo}
J.~Seo, Y.~Sung, G.~Lee, and D.~Kim, ``Pilot beam sequence design for channel
  estimation in millimeter-wave {MIMO} systems: A {POMDP} framework,'' in
  \emph{Proc. IEEE Signal Process. Adv. Wireless Commun.}, June 2015, pp.
  236--240.

\bibitem{rappaportMeasure}
G.~MacCartney and T.~Rappaport, ``73 {GH}z millimeter wave propagation
  measurements for outdoor urban mobile and backhaul communications in {N}ew
  {Y}ork {C}ity,'' in \emph{IEEE Int. Conf. on Commun. (ICC)}, June 2014, pp.
  4862--4867.

\bibitem{Akdeniz}
M.~Akdeniz, Y.~Liu, M.~Samimi, S.~Sun, S.~Rangan, T.~Rappaport, and E.~Erkip,
  ``Millimeter wave channel modeling and cellular capacity evaluation,''
  \emph{IEEE J. Select. Areas Commun.}, vol.~32, no.~6, pp. 1164--1179, June
  2014.

\bibitem{Compressed_Channel_Sensing}
W.~Bajwa, J.~Haupt, A.~Sayeed, and R.~Nowak, ``Compressed channel sensing: A
  new approach to estimating sparse multipath channels,'' \emph{Proc. IEEE},
  vol.~98, no.~6, pp. 1058--1076, June 2010.

\bibitem{IEEE_standard}
``Wireless medium access control (mac) and physical layer (phy) specifications
  for high rate wireless personal area networks ({WPAN})s, amendement 2:
  Millimeter-wave-based alternative physical layer extension, {IEEE} standard
  802.15.3c,'' Oct. 2009.

\bibitem{Tsang}
Y.~Tsang, A.~Poon, and S.~Addepalli, ``Coding the beams: Improving beamforming
  training in mmwave communication system,'' in \emph{Proc. IEEE Global
  Telecomm. Conf.}, Dec 2011, pp. 1--6.

\bibitem{Zhang_Kung}
X.~Zhang, A.~Molisch, and S.-Y. Kung, ``Variable-phase-shift-based
  {RF}-baseband codesign for mimo antenna selection,'' \emph{IEEE Trans. Signal
  Process.}, vol.~53, no.~11, pp. 4091--4103, Nov 2005.

\bibitem{xiao2016hierarchical}
Z.~Xiao, T.~He, P.~Xia, and X.-G. Xia, ``Hierarchical codebook design for
  beamforming training in millimeter-wave communication,'' \emph{IEEE Trans.
  Commun.}, vol.~15, no.~5, pp. 3380--3392, 2016.

\bibitem{xia2008multi}
P.~Xia, S.-K. Yong, J.~Oh, and C.~Ngo, ``Multi-stage iterative antenna training
  for millimeter wave communications,'' in \emph{Proc. IEEE Global Telecomm.
  Conf.}\hskip 1em plus 0.5em minus 0.4em\relax IEEE, 2008, pp. 1--6.

\bibitem{xiao2015iterative}
Z.~Xiao, X.-G. Xia, D.~Jin, and N.~Ge, ``Iterative eigenvalue decomposition and
  multipath-grouping tx/rx joint beamformings for millimeter-wave
  communications,'' \emph{IEEE Trans. Commun.}, vol.~14, no.~3, pp. 1595--1607,
  2015.

\bibitem{xiao2014iterative}
Z.~Xiao, L.~Bai, and J.~Choi, ``Iterative joint beamforming training with
  constant-amplitude phased arrays in millimeter-wave communications,''
  \emph{IEEE Commun. Letters}, vol.~18, no.~5, pp. 829--832, 2014.

\bibitem{xia2008practical}
P.~Xia, H.~Niu, J.~Oh, and C.~Ngo, ``Practical antenna training for millimeter
  wave mimo communication,'' in \emph{IEEE Vehicular Technology Conference},
  2008.

\bibitem{Sayeed_max}
A.~Sayeed and V.~Raghavan, ``The ideal {MIMO} channel: Maximizing capacity in
  sparse multipath with reconfigurable arrays,'' in \emph{Proc. ISIT}, July
  2006, pp. 1036--1040.

\bibitem{proakis}
J.~G. Proakis, \emph{Digital communications}.\hskip 1em plus 0.5em minus
  0.4em\relax New York, NY, USA: McGraw-Hill, 1995.

\bibitem{gallager2008circularly}
R.~G. Gallager, ``Circularly-symmetric gaussian random vectors,''
  \emph{preprint}, pp. 1--9, 2008.

\bibitem{bock1981marginal}
R.~D. Bock and M.~Aitkin, ``Marginal maximum likelihood estimation of item
  parameters: Application of an em algorithm,'' \emph{Psychometrika}, vol.~46,
  no.~4, pp. 443--459, 1981.

\bibitem{scharf1991statistical}
L.~L. Scharf, \emph{Statistical signal processing}.\hskip 1em plus 0.5em minus
  0.4em\relax Addison-Wesley Reading, MA, 1991, vol.~98.

\bibitem{stuber2011principles}
G.~L. St{\"u}ber, \emph{Principles of mobile communication}.\hskip 1em plus
  0.5em minus 0.4em\relax Springer Science \& Business Media, 2011.

\bibitem{shannon2001mathematical}
C.~E. Shannon, ``A mathematical theory of communication,'' \emph{Bell Syst.
  Tech. J.}, vol.~27, p. 379–423, 1948.

\bibitem{bai2014coverage}
T.~Bai, A.~Alkhateeb, and R.~Heath, ``Coverage and capacity of millimeter-wave
  cellular networks,'' \emph{IEEE Commun. Mag.}, vol.~52, no.~9, pp. 70--77,
  2014.

\end{thebibliography}

\end{document}